\documentclass{article}
\usepackage[utf8]{inputenc}

\usepackage[a4paper, top=1.25in, bottom=1.25in, left=1in, right=1in]{geometry}

\usepackage[font=small,labelfont=bf]{caption}

\usepackage{amsmath}
\usepackage{amsfonts}
\usepackage{bm}

\usepackage{graphicx}
\usepackage[font=small,labelfont=bf,labelsep=period]{caption}

\usepackage{pdfpages}

\makeatletter 
\renewcommand{\thefigure}{\@arabic\c@figure}
\makeatother

\usepackage{hyperref}
\hypersetup{colorlinks=true,citecolor=blue}

\usepackage[round, authoryear]{natbib}
\newcommand{\X}{\mathbf X}

\title{\textbf{Demixed principal component analysis of population activity in higher cortical areas reveals independent representation of task parameters}}
\author{Dmitry Kobak\textsuperscript{1,*}, Wieland Brendel\textsuperscript{1,2,*}, Christos Constantinidis\textsuperscript{3}, \\ Claudia E. Feierstein\textsuperscript{1},
Adam Kepecs\textsuperscript{4}, Zachary F. Mainen\textsuperscript{1}, \\ Ranulfo Romo\textsuperscript{5}, Xue-Lian Qi\textsuperscript{3}, Naoshige Uchida\textsuperscript{6}, and Christian K. Machens\textsuperscript{1} \\ \\
\textsuperscript{1}Champalimaud Centre for the Unknown, Lisbon, Portugal\\
\textsuperscript{2}{\'E}cole Normale Sup{\'e}rieure, Paris, France\\
\textsuperscript{3}Wake Forest University School of Medicine, Winston-Salem, NC, USA\\
\textsuperscript{4}Cold Spring Harbor Laboratory, NY, USA\\
\textsuperscript{5}Universidad Nacional Aut{\'o}noma de M{\'e}xico, Mexico\\
\textsuperscript{6}Harvard University, Cambridge, MA, USA\\ \\
\textsuperscript{*}Equal contribution}
\date{October 2014}

\begin{document}

\maketitle

\begin{abstract}
Neurons in higher cortical areas, such as the prefrontal cortex, are known to be tuned to a variety of sensory and motor variables. The resulting diversity of neural tuning often obscures the represented information. Here we introduce a novel dimensionality reduction technique, demixed principal component analysis (dPCA), which automatically discovers and highlights the essential features in complex population activities. We reanalyze population data from the prefrontal areas of rats and monkeys performing a variety of working memory and decision-making tasks. In each case, dPCA  summarizes the relevant features of the population response in a single figure. The population activity is decomposed into a few demixed components that capture most of the variance in the data and that highlight dynamic tuning of the population to various task parameters, such as stimuli, decisions, rewards, etc. Moreover, dPCA reveals strong, condition-independent components of the population activity that remain unnoticed with conventional approaches.
\end{abstract}

\section*{Introduction}

In many state of the art experiments, a subject, such as a rat or a monkey, performs a behavioral task while the activity of tens to hundreds of neurons in the animal's brain is monitored using electrophysiological or imaging techniques. The common goal of these studies is to relate the external task parameters, such as stimuli, rewards, or the animal's actions, to the internal neural activity, and to then draw conclusions about brain function. This approach has typically relied on the analysis of single neuron recordings. However, as soon as hundreds of neurons are taken into account, the complexity of the recorded data poses a fundamental challenge in itself. This problem has been particularly severe in higher-order areas such as the prefrontal cortex, where neural responses display a baffling heterogeneity, even if animals are carrying out rather simple tasks \citep{brody2003timing, machens2010demixing, mante2013context, cunningham2014dimensionality}.

Traditionally, this heterogeneity has often been ignored. In neurophysiological studies, it is common practice to pre-select cells based on particular criteria, such as responsiveness to the same stimulus, and to then average the firing rates of the pre-selected cells. This practice eliminates much of the richness of single-cell activities, similar to imaging techniques with low spatial resolution, such as MEG, EEG, or fMRI. While population averages can identify the information that higher-order areas process, they ignore how exactly that information is represented on the neuronal level \citep{wohrer2013population}. Indeed, most neurons in higher cortical areas will typically encode several task parameters simultaneously, and therefore display what has been termed “mixed selectivity” \citep{rigotti2013importance}.

Instead of looking at single neurons and selecting from or averaging over a population of neurons, neural population recordings can be analyzed using dimensionality reduction methods \citep[for a review, see][]{cunningham2014dimensionality}. In recent years, several such methods have been developed that are specifically targeted to electrophysiological data, taking into account the binary nature of spike trains \citep{yu2009gaussian, pfau2013robust}, or the dynamical properties of the population response \citep{buesing2012spectral, churchland2012neural}. However, these approaches reduce the dimensionality of the data without taking task parameters, i.e., sensory and motor variables controlled or monitored by the experimenter, into account. Consequently, mixed selectivity remains in the data even after the dimensionality reduction step, impeding interpretation of the results.

A few recent studies have sought to solve these problems by developing methods that reduce the dimensionality of the data in a way that is informed by the task parameters \citep{machens2010demixing, machens2010functional, brendel2011demixed, mante2013context}. One possibility is to adopt a parametric approach, i.e. to assume a specific (e.g. linear) dependency of the firing rates on the task parameters, and then use regression to construct variables that demix the task components \citep{mante2013context}. While this method can help to sort out the mixed selectivities, it still runs the risk of missing important structures in the data if the neural activities do not conform to the dependency assumptions (e.g. because of nonlinearities).

Here, we follow an alternative route by developing an unbiased dimensionality reduction technique that fulfills two constraints. It aims to find a decomposition of the data into latent components that (a) are easily interpretable with respect to the experimentally controlled and monitored task parameters; and (b) preserve the original data as much as possible, ensuring that no valuable information is thrown away. Our method, which we term demixed principal component analysis (dPCA), improves our earlier methodological work \citep{machens2010demixing, machens2010functional, brendel2011demixed} by eliminating unnecessary orthogonality constraints on the decomposition. In contrast to several recently suggested algorithms for decomposing firing rates of individual neurons into demixed parts \citep{pagan2014quantifying, park2014encoding}, our focus is on dimensionality reduction.

There is no a priori guarantee that neural population activities can be linearly demixed into latent variables that reflect individual task parameters. Nevertheless, we applied dPCA to spike train recordings from monkey prefrontal cortex (PFC) \citep{romo1999neuronal, qi2011changes} and from rat orbitofrontal cortex (OFC) \citep{feierstein2006representation, kepecs2008neural}, and obtained remarkably successful demixing. In each case, dPCA automatically summarizes all the important, previously described features of the population activity in a single figure. Importantly, our method provides an easy visual comparison of complex population data across different tasks and brain areas, which allows us to highlight both similarities and differences in the neural activities. Demixed PCA also reveals several hitherto largely ignored features of population data: (1) most of the activity in these datasets is not related to any of the controlled task parameters, but depends only on time (``condition-independent activity''); (2) all measured task parameters can be extracted with essentially orthogonal linear readouts; and (3) task-relevant information is shifted around in the neural population space, moving from one component to another during the course of the trial.

\section*{Results}

\subsection*{Demixed Principal Component Analysis (dPCA)}

\begin{figure}
\includegraphics[width=1\linewidth]{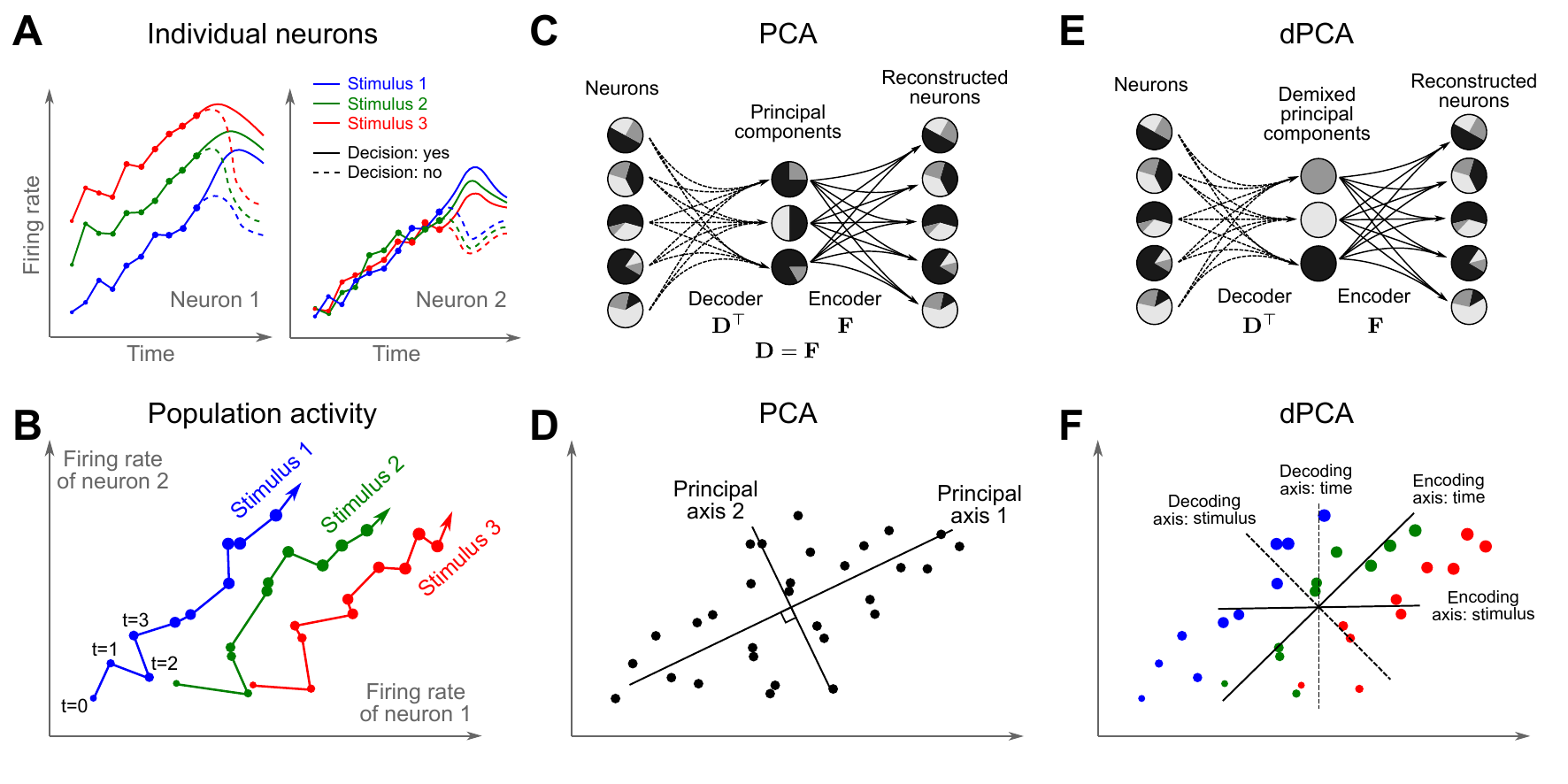}
\caption{Standard and demixed principal component analyses. \textbf{(A)} Time-varying firing rates of two neurons in a simple stimulus-decision task. Each subplot shows one neuron. Each line shows one condition, with colour coding the stimulus and solid/dashed line coding the binary decision. \textbf{(B)} Sketch of the population firing rate. At any given moment in time, the population firing rate of $N$ neurons is represented by a point in an $N$-dimensional space; here $N=2$. Each trial is represented by a trajectory in this space. Different experimental conditions (e.g. different stimuli) lead to different average trajectories. Here, colours indicate the different stimuli and dot sizes represent the time points. The decision period is not shown for visual clarity. \textbf{(C)} Principal component analysis (PCA). The firing rates of the neurons are linearly transformed (with a “decoder”) into a few latent variables (known as principal components). A second linear transformation (“encoder”) can reconstruct the original firing rates from these principal components. PCA finds the decoder/encoder transformations that minimize the reconstruction error. Shades of grey  inside each neuron/component show the proportion of variance due to the various task parameters (e.g. stimulus, decision, and time), illustrating the “mixed selectivity” of both, neurons and principal components. Variance terms due to parameter interactions are omitted for clarity. \textbf{(D)} Geometric intuition behind PCA. Same set of points as in (B). The lines show the first two principal axes, the longer one corresponds to the principal component explaining more variance (the first PC). These two axes form both the decoder and the encoder, which in case of PCA are identical. \textbf{(E)} Demixed principal component analysis (dPCA). As in PCA, the firing rates are compressed and decompressed through two linear transformations. However, here the transformations are found by both minimizing the reconstruction error and enforcing a demixing constraint on the latent variables. \textbf{(F)} Geometric intuition behind dPCA. Same set of points as in (B,D), now labeled with two parameters: stimulus (colour), and time (dot size); decision is ignored for simplicity. The dashed lines show the decoding axes, i.e., the axes on which points are projected such that the task-parameter dependencies are demixed. Solid lines show the encoding axes, i.e., the axes along which the demixed principal components need to be re-expanded in order to reconstruct the original data. }
\label{fig:intro}
\end{figure}

We illustrate our method with a toy example (Figure 1). Consider a standard experimental paradigm in which a subject is first presented with a stimulus and then reports a binary decision. The firing rates of two simulated neurons are shown in Figure 1A. The first neuron's firing rate changes with time, with stimulus (at the beginning of the trial), and with the animal's decision (at the end of the trial). The second neuron's firing rate also changes with time, but only depends on the subject's decision (not on the stimulus). As time progresses within a trial, the joint activity of the two neurons traces out a trajectory in the space of firing rates (Figure 1B; decision not shown for visual clarity). More generally, firing rates of a population of $N$ neurons at any moment in time is represented by a point in an $N$-dimensional space.

One of the standard approaches to reduce the dimensionality of complex, high-dimensional datasets is principal component analysis (PCA). In PCA, the original data (here, firing rates of the neurons) are linearly transformed into several latent variables, called principal components. Each principal component is therefore a linear combination, or weighted average, of the actual neurons. We interpret these principal components as linear “read-outs” from the full population (Figure 1C, “decoder” $\mathbf D$). This transformation can be viewed as a compression step, as the number of informative latent variables is usually much smaller than the original dimensionality. The resulting principal components can then be de-compressed with another linear mapping (“encoder” $\mathbf F$), approximately reconstructing the original data. Geometrically, when applied to a cloud of data points (Figure 1D), PCA constructs a set of directions (principal axes) that consecutively maximize the variance of the projections; these axes form both the decoder and the encoder.

More precisely, given a data matrix $\X$, in which each row contains the firing rates of one neuron for all task conditions, PCA finds decoder and encoder that minimize the loss function
$$L_\mathrm{PCA}  = \|\X - \mathbf F \mathbf D \X\|^2$$ 
under the constraint that the principal axes are normalized and orthogonal, and therefore $\mathbf D= \mathbf F^\top$ \citep[section 14.5]{hastie2009elements}. However, applying PCA to a neural dataset like the one sketched in Figure 1A generally results in principal components exhibiting mixed selectivity: many components will vary with time, stimulus, and decision (see examples below). Indeed, information about the task parameters does not enter the loss function.

To circumvent this problem, we sought to develop a new method by introducing an additional constraint: the latent variables should not only compress the data optimally, but also demix the selectivities (Figure 1E). As in PCA, compression is achieved with a linear mapping $\mathbf D$ and decompression with a linear mapping $\mathbf F$. Unlike PCA, the decoding axes are not constrained to be orthogonal to each other and may have to be non-orthogonal, in order to comply with demixing. The geometrical intuition is presented in Figure 1F. Here, the same cloud of data points as in Figures 1B and 1D has labels corresponding to time and stimulus. When we project the data onto the time-decoding axis, information about time is preserved while information about stimulus is lost, and vice versa for the stimulus-decoding axis. Hence, projections on the decoding axes demix the selectivities. Since the data have been projected onto a non-orthogonal (affine) coordinate system, the optimal reconstruction of the data requires a de-compression along a different set of axes, the encoding axes (which implies that $\mathbf D \approx \mathbf F^+$, the pseudo-inverse of $\mathbf F$, see Methods).

Given the novel demixing constraint, we term our method demixed PCA (dPCA), and provide a detailed mathematical account in the Methods. Briefly, in the toy example of Figure 1, dPCA first splits the data matrix into a stimulus-varying part $\X_s$ and a time-varying part $\X_t$, so that $\X \approx \X_t + \X_s$, and then finds separate decoder and encoder matrices for stimulus and time by minimizing the loss function 
$$L_\mathrm{dPCA} = \|\X_s - \mathbf F_s \mathbf D_s \X\|^2 + \|\X_t - \mathbf F_t \mathbf D_t \X\|^2.$$
See Methods for a more general case.

\subsection*{Somatosensory working memory task in monkey PFC}

\begin{figure}
\includegraphics[width=1\linewidth]{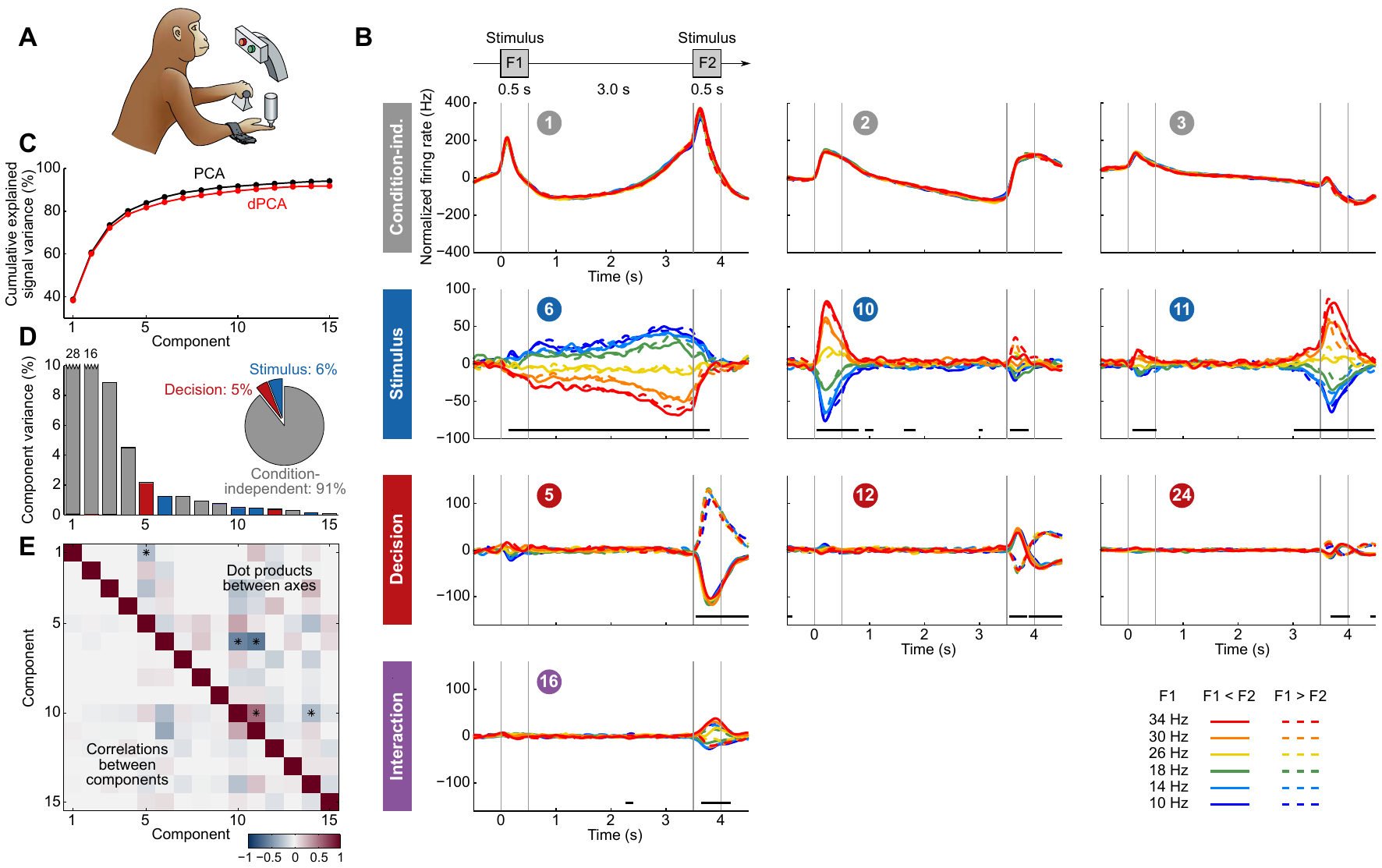}
\caption{Demixed PCA applied to recordings from monkey PFC during a somatosensory working memory task \citep{romo1999neuronal}. \textbf{(A)} Cartoon of the paradigm, adapted from \citep{romo2003flutter}. \textbf{(B)} Demixed principal components. Top row: first three condition-independent components; second row: first three stimulus components; third row: first three decision components; last row: first stimulus/decision interaction component. In each subplot there are 12 lines corresponding to 12 conditions (see legend). Thick black lines show time intervals during which the respective task parameters can be reliably extracted from single-trial activity (see Methods). Note that the vertical scale differs across rows. \textbf{(C)} Cumulative signal variance explained by PCA (black) and dPCA (red). Demixed PCA explains almost the same amount of variance as standard PCA. \textbf{(D)} Variance of the individual demixed principal components. Each bar shows the proportion of total variance, and is composed out of four stacked bars of different colour: gray for condition-independent variance, blue for stimulus variance, red for decision variance, and purple for variance due to stimulus-decision interactions. Each bar appears to be single-colored, which signifies nearly perfect demixing. Pie chart shows how the total signal variance is split between parameters. \textbf{(E)} Upper-right triangle shows dot products between all pairs of the first 15 demixed principal axes. Stars mark the pairs that are significantly and robustly non-orthogonal (see Methods). Bottom-left triangle shows correlations between all pairs of the first 15 demixed principal components. Most of the correlations are close to zero.}
\label{fig:romo}
\end{figure}

We first applied dPCA to recordings from the PFC of monkeys performing a somatosensory working memory task \citep{romo1999neuronal, brody2003timing}. In this task, monkeys were required to discriminate two vibratory stimuli presented to the fingertip. The stimuli were separated by a 3~s delay, and the monkeys had to report which stimulus had a higher frequency by pressing one of two available buttons (Figure 2A). Here, we focus on the activity of 832 prefrontal neurons from two animals, see Methods. For each neuron, we obtained the average time-dependent firing rate (also known as peri-stimulus time histogram, PSTH) in 12 conditions: 6 values of stimulus frequency F1, each paired with 2 possible decisions of the monkey. We only analyzed correct trials, so that the decision of the monkey always coincided with the actual trial type (F2\,$>$\,F1 or F1\,$>$\,F2).

As is typical for PFC, each neuron has a distinct response pattern and many neurons show mixed selectivity (Figure S1). Several previous studies have sought to make sense of these heterogeneous response patterns by separately analyzing different task periods, such as the stimulation and delay periods \citep{romo1999neuronal, brody2003timing, machens2010functional, barak2010neuronal}, the decision period \citep{jun2010heterogenous}, or both \citep{hernandez2010decoding}. With dPCA, however, we can summarize the main features of the neural activity across the whole trial in a single figure (Figure 2B). Just as in PCA, we can think of these demixed principal components as the “building blocks” of the observed neural activity, in that the activity of each single neuron is a linear combination (weighted average) of these components. These building blocks come in four distinct categories: some are condition-independent (Figure 2B, top row); some depend only on stimulus F1 (second row); some depend only on decision (third row); and some depend on stimulus and decision together (bottom row). The components in the first three categories demix the parameter dependencies, which is exactly what dPCA aimed for. We stress that applying standard PCA to the same dataset leads to components that strongly exhibit mixed selectivity (Figure S2; see Methods for quantification).
 
What can we learn about the population activity from the dPCA decomposition? First, we find that information about the stimulus and the decision can be fully demixed at the population level, even though it is mixed at the level of individual neurons. Importantly, the demixing procedure is linear, so that all components could in principle be retrieved by the nervous system through synaptic weighting and dendritic integration of the neurons' firing rates. Note that our ability to linearly demix these different types of information is a non-trivial feature of the data.

Second, we see that most of the variance of the neural activity is captured by the condition-independent components (Figure 2B, top row) that together amount to $\sim$90\% of the signal variance (see the pie chart in Figure 2D and Methods). These components capture the temporal modulations of the neural activity throughout the trial, irrespective of the task condition. Their striking dominance in the data may come as a surprise, as such condition-independent components are usually not analyzed or shown. Previously, single-cell activity in the delay period related to these components has often been dubbed “ramping activity” or “climbing activity” \citep{durstewitz2004neural, rainer1999prospective, komura2001retrospective, brody2003timing, janssen2005representation, machens2010functional}; however, analysis of the full trial here shows that the temporal modulations of the neural activity are far more complex and varied. Indeed, they spread out over many components. The origin of these components can only be conjectured, see Discussion. 
 
Third, our analysis also captures the major findings previously obtained with these data: the persistence of the F1 tuning during the delay period --- component \#6 \citep{romo1999neuronal,machens2005flexible}, the temporal dynamics of short-term memory --- components \#\#6, 10, 11 \citep{brody2003timing, machens2010functional, barak2010neuronal}, the ramping activities in the delay period --- components \#\#1--3, 6 \citep{singh2006higher, machens2010functional}; and pronounced decision-related activities --- component \#5 \citep{jun2010heterogenous}. We note that the decision components resemble derivatives of each other; these higher-order derivatives likely arise due to slight variations in the timing of responses across neurons (Figure S3).

Fourth, we observe that the three stimulus components show the same monotonic (“rainbow”) tuning but occupy three distinct time periods: one is active during the S1 period (\#10), one during the delay period (\#6), and one during the S2 period (\#11). Information about the stimulus is therefore shifted around in the high-dimensional firing rate space. Indeed, if the stimulus were always encoded in the same subspace, there would only be a single stimulus component. In other words, if we perform dPCA analysis in a short sliding time window, then the main stimulus axis will rotate with time instead of remaining constant.

We furthermore note several important technical points. First, the overall variance explained by the dPCA components (Figure 2C, red line) is very close to the overall variance captured by the PCA components (Figure 2C, black line). This means that by imposing the demixing constraint we did not lose much of the variance, and so the population activity is accurately represented by the obtained dPCA components (Figure 2B). Second, as noted above, the demixed principal axes are not constrained to be orthogonal. Nonetheless, most of them are quite close to orthogonal (Figure 2E, upper-right triangle). There are a few notable exceptions, which we revisit in the Discussion section. Third, pairwise correlations between components are all close to zero (Figure 2E, lower-left triangle), as should be expected as the components are considered to represent independent signals.

\subsection*{Visuospatial working memory task in monkey PFC}

We next applied dPCA to recordings from the PFC of monkeys performing a visuospatial working memory task \citep{qi2011changes, meyer2011stimulus, qi2012neural}. In this task, monkeys first fixated at a small white square at the centre of a screen, when a square S1 appeared in one of the eight locations around the centre (Figure 3A). After a 1.5 s delay, a second square S2 appeared in either the same (“match”) or the opposite (“non-match”) location. Following another 1.5 s delay, a green and a blue choice targets appeared in locations orthogonal to the earlier presented stimuli. Monkeys had to saccade to the green target to report a match condition, and to the blue one to report a non-match.

\begin{figure}
\includegraphics[width=1\linewidth]{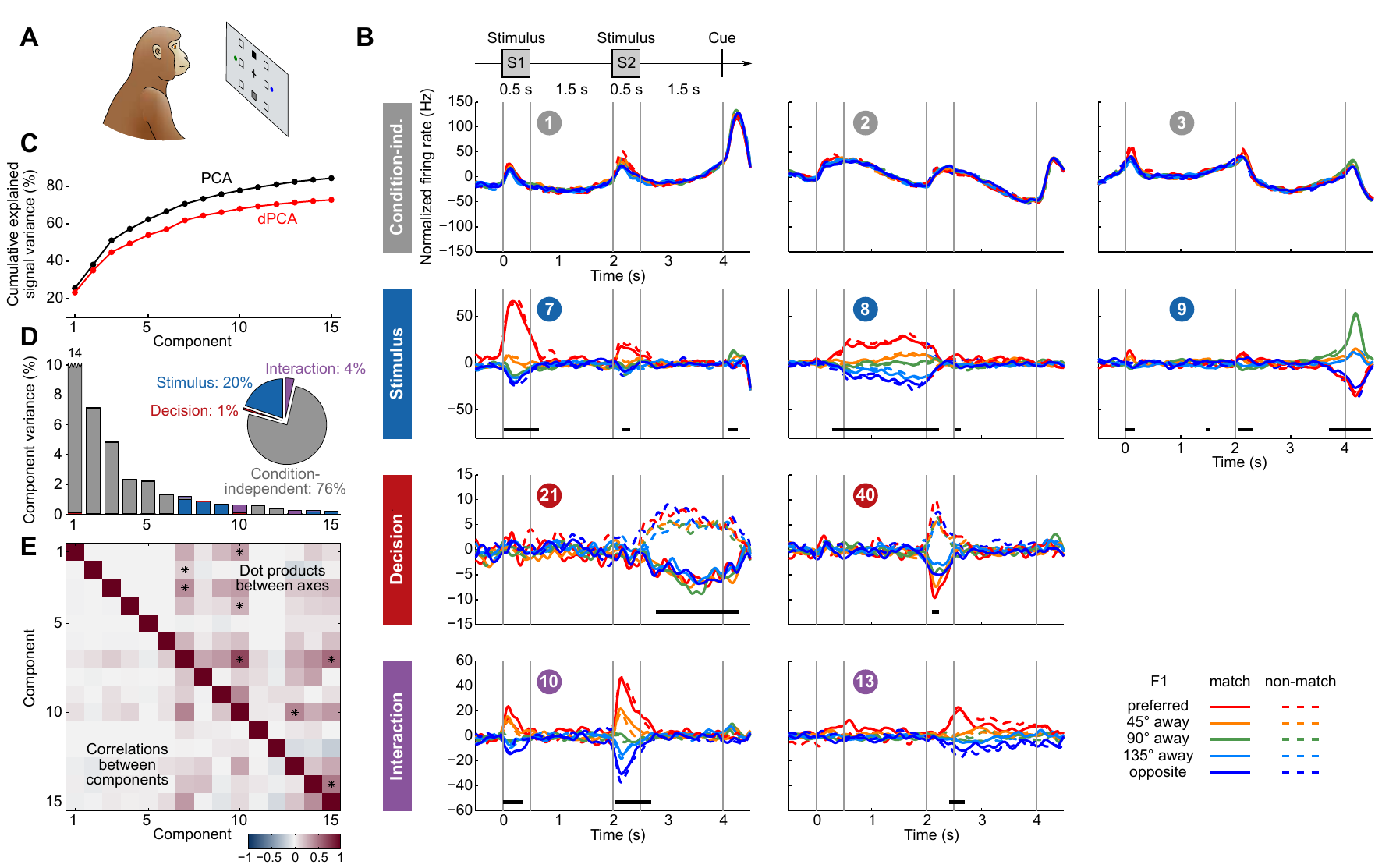}
\caption{Demixed PCA applied to recordings from monkey PFC during a visuospatial working memory task \citep{qi2011changes}. Same format as Figure 2. \textbf{(A)} Cartoon of the paradigm, adapted from \citep{romo2003flutter}. \textbf{(B)} Demixed principal components. In each subplot there are 10 lines corresponding to 10 conditions (see legend). Colour corresponds to the position of the last shown stimulus (first stimulus for $t<2$ s, second stimulus for $t>2$ s). In non-match conditions (dashed lines) the colour changes at $t=2$ s. Solid lines correspond to match conditions and do not change colours. \textbf{(C)} Cumulative signal variance explained by PCA and dPCA components. \textbf{(D)} Variance of the individual demixed principal components. Pie chart shows how the total signal variance is split between parameters. \textbf{(E)} Upper-right triangle shows dot products between all pairs of the first 15 demixed principal axes, bottom-left triangle shows correlations between all pairs of the first 15 demixed principal components.}
\label{fig:const}
\end{figure}

We analyzed the activity of 956 neurons recorded in the lateral PFC of two monkeys performing this task. Proceeding exactly as before, we obtained the average time-dependent firing rate of each neuron for each condition. Following the original studies, we eliminated the trivial rotational symmetry of the task by collapsing the eight possible stimulus locations into five locations that are defined with respect to the preferred direction of each neuron (0$^\circ$, 45$^\circ$, 90$^\circ$, 135$^\circ$, or 180$^\circ$ away from the preferred direction, see Methods). As a consequence, we obtained 10 conditions: 5 possible stimulus locations, each paired with 2 possible decisions of the monkey. We again only analyzed correct trials, so that the decision of the monkey always coincided with the actual trial type.

The dPCA results are shown in Figure 3. Just as before, the components fall into four categories: they can be condition-independent (Figure 3B, top row); dependent only on stimulus location (second row); only on decision (third row); or dependent on stimulus-decision interactions (bottom row).

We note several similarities to the somatosensory working memory task. First, stimulus and decision can be separated at the population level just as easily as before, despite being intermingled at the single-neuron level. Second, most of the variance in the neural activity is again captured by the condition-independent components (Figure 3B, top row, and 3D, pie chart).  Third, all stimulus components are active in different time periods, and the same is true for the decision, indicating rotation of the stimulus and decision representations in the firing rate space. Fourth, and maybe most surprisingly, we find that the overall structure of population activity up to the second stimulus (S2) is almost identical to that found in the somatosensory task. In both cases, the leading condition-independent components are quite similar (compare Figure 2B with Figure 3B). Furthermore, in both tasks we find a stimulus component with strong activity during the F1/S1 period, a stimulus component with persistent activity during the delay period, and a decision component with activity during the F2/S2 period. 

One notable difference between Figures 2 and 3 is the presence of strong interaction components in Figure 3B. However, interaction components in Figure 3B are in fact stimulus components in disguise. In match trials, S2 and S1 appear at the same location, and in non-match trials at opposite locations. Information about S2 is therefore given by a non-linear function of stimulus S1 and the trial type (i.e. decision), which is here captured by the interaction components. 

The differences in the population activity between these two tasks seem to stem mostly from the specifics of the tasks, such as the existence of a second delay period in the visuospatial working memory task. There are also differences in the overall amount of power allocated to different components, with the stimulus (and interaction) components dominating in the visuospatial working memory task, whereas stimulus and decision components are more equally represented in the somatosensory task. However, the overall structure of the data is surprisingly similar.

Here again, our analysis summarizes previous findings obtained with this dataset. For instance, the first and the second decision components show tuning to the match/non-match decision in the delay period between S2 and the cue (first component) and during the S2 period (second component). Using these components as fixed linear decoders, we achieve cross-validated single-trial classification accuracy of match vs. non-match of $\sim$75\% for $t>2$ (Figure S4), which is approximately equal to the state-of-the-art classification performance reported previously \citep{meyers2012incorporation}.

Constantinidis et al. have also recorded population activity in PFC before starting the training (both S1 and S2 stimuli were presented exactly as above, but there were no cues displayed and no decision required). When analyzing this pre-training population activity with dPCA, the first stimulus and the first interaction components come out close to the ones shown on Figure 3, but there are no decision and no “memory” components present (Figure S5), in line with previous findings \citep{meyers2012incorporation}. These task-specific components appear in the population activity only after extensive training.

\subsection*{Olfactory discrimination task in rat OFC}

Next, we applied dPCA to recordings from the OFC of rats performing an odour discrimination task \citep{feierstein2006representation}. This behavioral task differs in two crucial aspects from the previously considered tasks: it requires no active storage of a stimulus, and it is self-paced. To start a trial, rats entered an odour port, which triggered delivery of an odour with a random delay of 0.2--0.5 s. Each odour was uniquely associated with one of the two available water ports, located to the left and to the right from the odour port (Figure 4A). Rats could sample the odour for as long as they wanted up to 1 s, and then had to move to one of the water ports. If they chose the correct water port, reward was delivered following an anticipation period of random length (0.2--0.5 s).

\begin{figure}[t]
\includegraphics[width=1\linewidth]{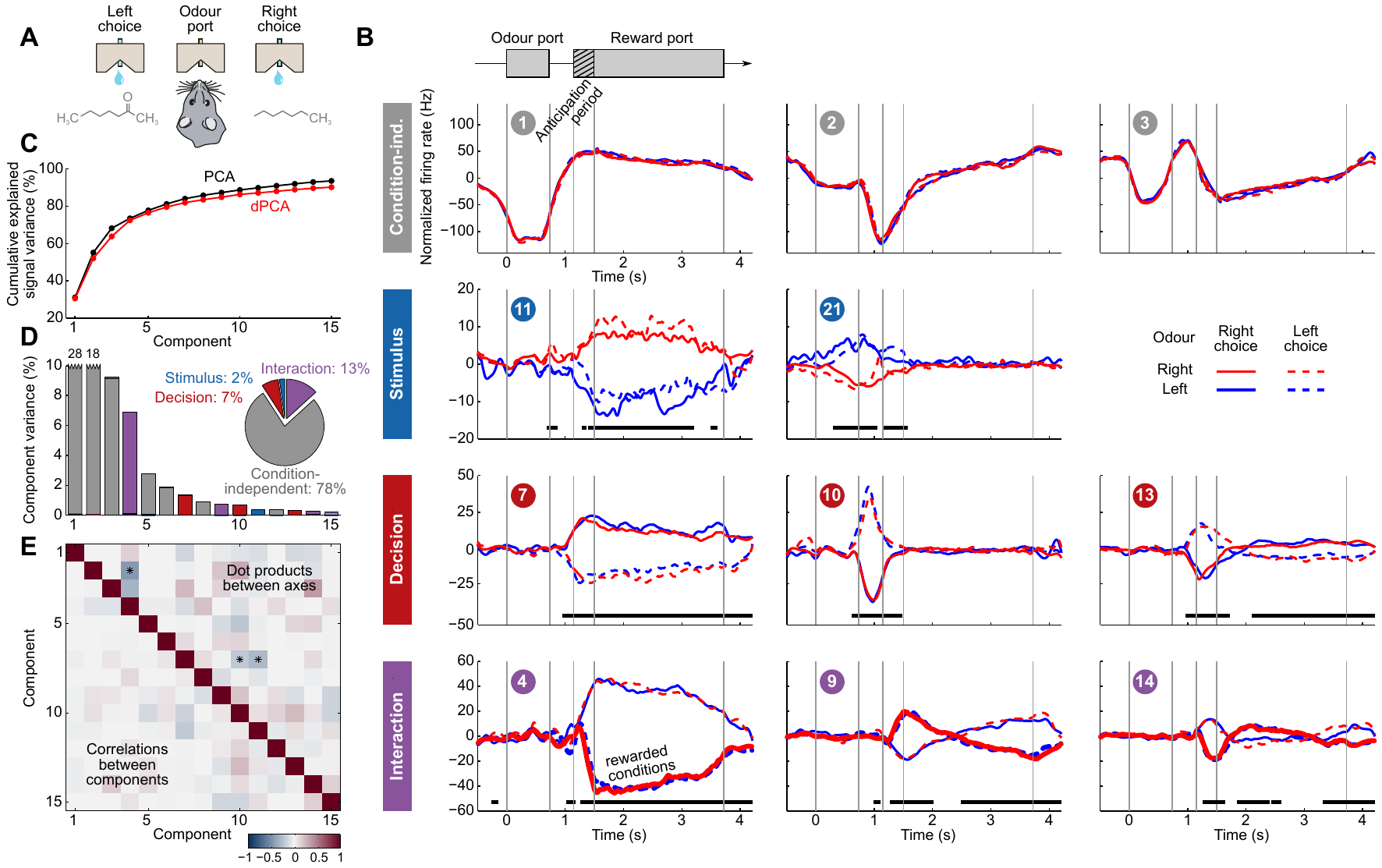}
\caption{Demixed PCA applied to recordings from rat OFC during an olfactory discrimination task \citep{feierstein2006representation}. Same format as Figure 2. \textbf{(A)} Cartoon of the paradigm, adapted from \citep{wang2013dorsomedial}. \textbf{(B)} Each subplot shows one demixed principal component. In each subplot there are 4 lines corresponding to 4 conditions (see legend). Two out of these four conditions were rewarded and are shown by thick lines. \textbf{(C)} Cumulative variance explained by PCA and dPCA components. \textbf{(D)} Variance of the individual demixed principal components. Pie chart shows how the total signal variance is split between parameters. \textbf{(E)} Upper-right triangle shows dot products between all pairs of the first 15 demixed principal axes, bottom-left triangle shows correlations between all pairs of the first 15 demixed principal components.}
\label{fig:clau}
\end{figure}

We analyzed the activity of 437 neurons recorded in five rats in 4 conditions: 2 stimuli (left and right) each paired with 2 decisions (left and right). Two of these conditions correspond to correct (rewarded) trials, and two correspond to error (unrewarded) trials. Since the task was self-paced, each trial had a different length; in order to align events across trials, we restretched the firing rates in each trial (see Methods). Alignment methods without restretching led to similar results (Figure S6).

Just as neurons from monkey PFC, neurons in rat OFC exhibit diverse firing patterns and mixed selectivity \citep{feierstein2006representation}. Nonetheless, dPCA is able to demix the population activity, resulting in the condition-independent, stimulus, decision, and interaction components (Figure 4). In this dataset, interaction components separate rewarded and unrewarded conditions (thick and thin lines on Figure 4B, bottom row), i.e., correspond to neurons tuned either to reward, or to the absence of reward.

We note several similarities to the monkey PFC data. The largest part of the total variance is again due to the condition-independent components (over 60\% of the total variance falls into the first three components). Also, for both stimulus and decision, the main components are localized in distinct time periods, meaning that information about the respective parameters is shifted around in firing rate space. 

The overall pattern of neural tuning across task epochs that we present here agrees with the findings of the original study \citep{feierstein2006representation}. Interaction components are by far the most prominent among all the condition-dependent components, corresponding to the observation that many neurons are tuned to presence/absence of reward. Decision components come next, with a caveat that decision information may also reflect the rat's movement direction and/or position, as was pointed out previously \citep{feierstein2006representation}. Stimulus components are less prominent, but nevertheless show clear stimulus tuning, demonstrating that even in error trials there is reliable information about stimulus identity in the population activity.

Finally, we note that the first interaction component (\#4) shows significant tuning to reward already in the anticipation period. In other words, neurons tuned to presence/absence of reward start firing before the reward delivery (or, on error trials, before the reward could have been delivered). We return to this observation in the next section.

\subsection*{Olfactory categorization task in rat OFC}

\citet{kepecs2008neural} extended the experiment of \citet{feierstein2006representation} by using odour mixtures instead of pure odours, thereby varying the difficulty of each trial \citep{uchida2003speed, kepecs2008neural}. In each trial, rats experienced mixtures of two fixed odours with different proportions (Figure 5A). Left choices were rewarded if the proportion of the “left” odour was above 50\%, and right choices otherwise. Furthermore, the waiting time until reward delivery (anticipation period) was increased to 0.3--2 s.

\begin{figure}[t]
\includegraphics[width=1\linewidth]{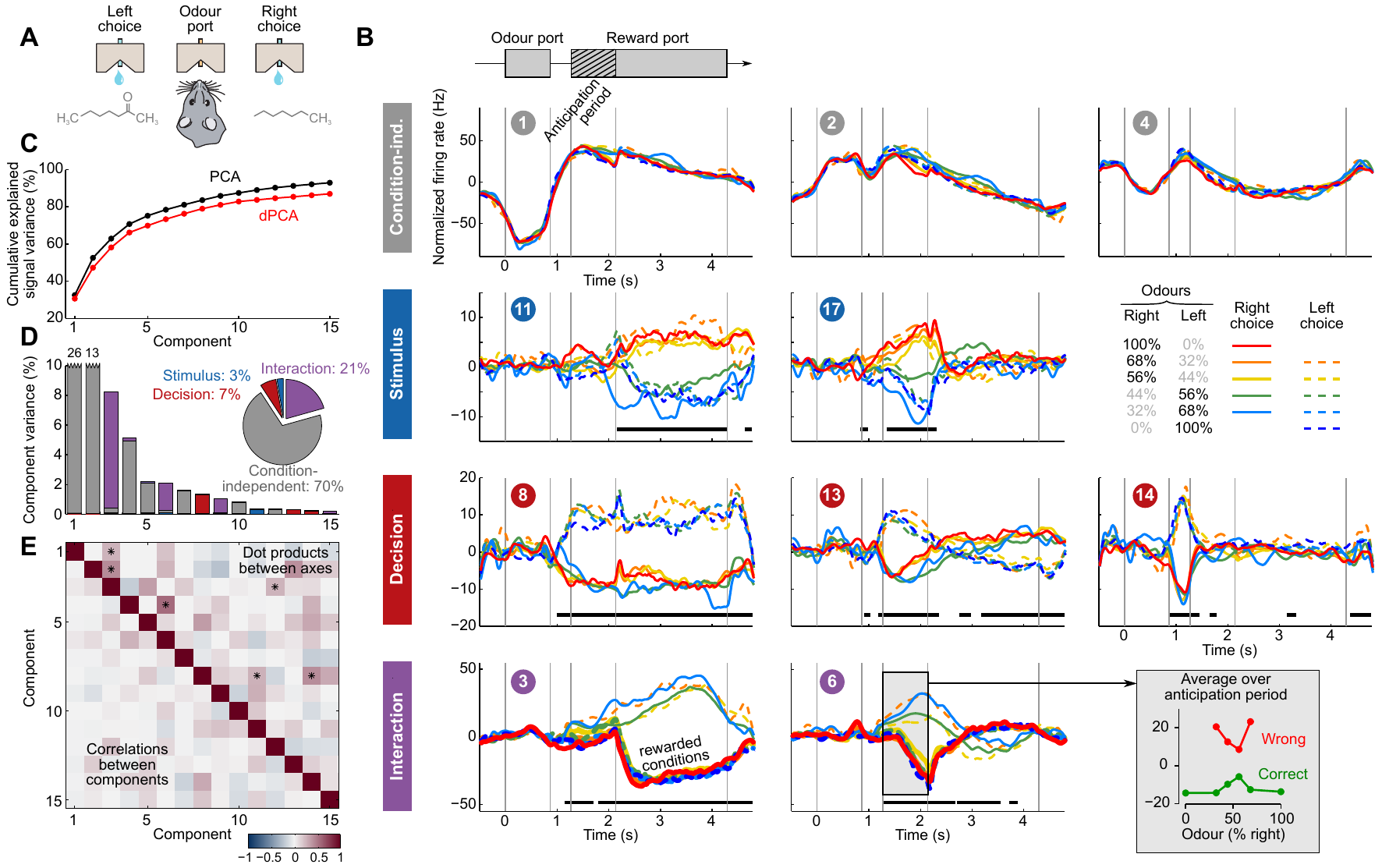}
\caption{Demixed PCA applied to recordings from rat OFC during an olfactory categorization task \citep{kepecs2008neural}. Same format as Figure 2. \textbf{(A)} Cartoon of the paradigm, adapted from \citep{wang2013dorsomedial}. \textbf{(B)} Each subplot shows one demixed principal component. In each subplot there are 10 lines corresponding to 10 conditions (see legend). Six out of these 10 conditions were rewarded and are shown with thick lines; note that the pure left (red) and the pure right (blue) odours did not have error trials. Inset shows mean rate of the second interaction component during the anticipation period. \textbf{(C)} Cumulative variance explained by PCA and dPCA components. \textbf{(D)} Variance of the individual demixed principal components. Pie chart shows how the total signal variance is split between parameters. \textbf{(E)} Upper-right triangle shows dot products between all pairs of the first 15 demixed principal axes, bottom-left triangle shows correlations between all pairs of the first 15 demixed principal components. }
\label{fig:adam}
\end{figure}

We analyzed the activity of 214 OFC neurons from three rats recorded in 8 conditions, corresponding to 4 odour mixtures, each paired with 2 decisions (left and right). During the presentation of pure odours (100\% right and 100\% left) rats made essentially no mistakes, and so we had to exclude these data from the dPCA computations (which require that all parameter combinations are present, see Methods). Nevertheless, we displayed these additional 2 conditions in Figure 5.

The dPCA components shown in Figure 5B are similar to those presented above in Figure 4B. The condition-independent components capture most of the total variance; the interaction components (corresponding to the reward) are most prominent among the condition-dependent ones; the decision components are somewhat weaker and show tuning to the rat's decision, starting from the odour port exit and throughout the rest of the trial; and the stimulus components are even weaker, but again reliable. 

Here again, some of the interaction components (especially the second one, \#6) show strong tuning already during the anticipation period, i.e. before the actual reward delivery. The inset in Figure 5B shows the mean value of the component \#6 during the anticipation period, separating correct (green) and incorrect (red) trials for each stimulus. The characteristic U-shape for the error trials and the inverted U-shape for the correct trials agrees well with the predicted value of the rat’s uncertainty in each condition \citep{kepecs2008neural}. Accordingly, this component can be interpreted as corresponding to the rat's uncertainty or confidence about its own choice, confirming the results of \citet{kepecs2008neural}. In summary, both the main features of this dataset, as well as some of the subtleties that have been pointed out before \citep{kepecs2008neural} are picked up and reproduced by dPCA.

\subsection*{Distribution of components in the neural population}

All dPCA components in each of the data sets are distributed across the whole neural population. For each component and each neuron, the corresponding encoder weight shows how much this particular component is exhibited by this particular neuron. For each component, the distribution of weights is strongly unimodal and centred at zero (Figure 6). In other words, there are no distinct sub-populations of neurons predominantly expressing a particular component; rather, each individual neuron can be visualized as a random linear combination of these components. We confirmed this observation by applying a recently developed clustering algorithm \citep{rodriguez2014clustering} to the population of neurons in the 15-dimensional space of dPC weights. In all cases, the algorithm found only one cluster (Figure S7; see Figure S8 for an analysis with Gaussian mixture models).

\begin{figure}[h]
\centering
\includegraphics[width=0.5\linewidth]{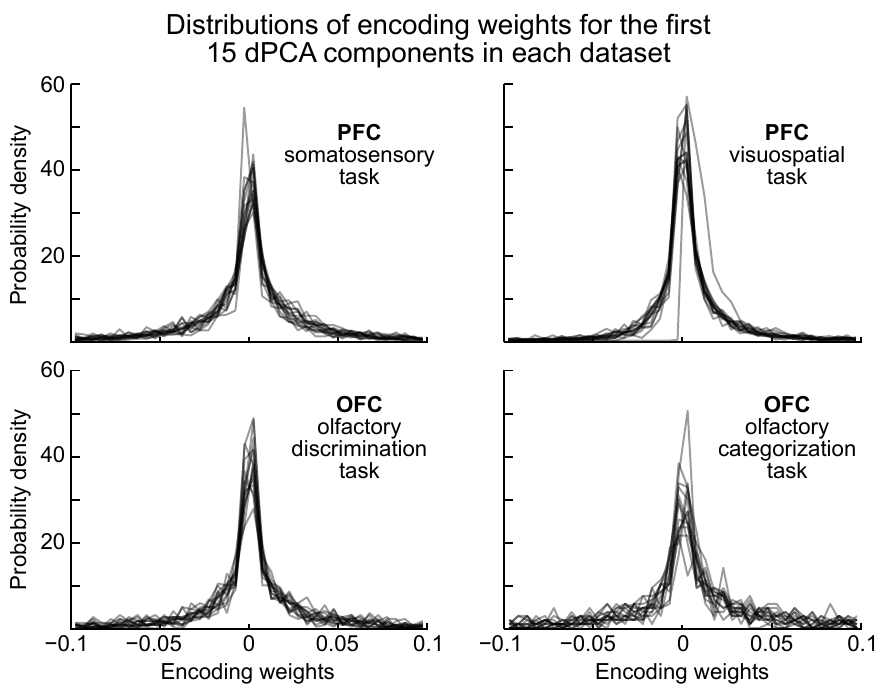}
\caption{Distribution of encoder weights for the leading dPCA components in the neural population. Each subplot shows 15 probability density curves, one curve per component (bin width 0.005).}
\label{fig:weights}
\end{figure}

\section*{Discussion}

The complexity of responses in higher-order areas such as the prefrontal or orbitofrontal cortices has plagued researchers for quite some time. Here we have introduced a new dimensionality reduction method (demixed PCA) that is specifically targeted to resolve these difficulties. Our method summarizes the data by extracting a set of latent components from the single-cell activities. The critical advantage of dPCA, compared to standard methods such as PCA or factor analysis (FA), is that individual components do not show mixed selectivity, but are instead “demixed”. This demixing can greatly simplify exploration and interpretation of neural data. Indeed, in all cases presented here, all the major aspects of the population activity that had previously been reported are directly visible on the dPCA summary figure.

In the following discussion we first compare dPCA with various alternative approaches to the analysis of high-dimensional neural datasets, and then recapitulate what dPCA teaches us about neural activity in higher-order areas.

\subsection*{Demixed PCA in comparison with other statistical methods}

\subsubsection*{I. Percentage of significantly responding neurons}

The most conventional and wide-spread method of analysis is to count the number of neurons that show significant response to a particular parameter of interest in a particular time period, and to then report the resulting percentage. This method was used in all of the original publications whose data we re-analyzed here: \citet{romo1999neuronal} showed that 65\% of the neurons (from those showing any task-related responses) exhibited monotonic tuning to the stimulation frequency during the delay period. \citet{meyer2011stimulus} reported that the percentage of neurons with increased firing rate in the delay period increased after training from 15--21\% to 27\%, and the percentage of neurons tuned to match/non-match difference increased from 11\% to 21\%. \citet{feierstein2006representation} found that 29\% of the neurons were significantly tuned to reward, and 41\% to the rat’s decision. Similarly, \citet{kepecs2008neural} found that 42\% of the neurons were significantly tuned to reward and observed that in the anticipation period these neurons demonstrated tuning to the expected uncertainty level. Note that in all of these cases our conclusions based on the dPCA analysis qualitatively agree with these previous findings.

Even though the conventional approach is perfectly sound, it does have several important limitations that dPCA is free of: 
\begin{enumerate}
\item The conventional analysis focuses on a particular time bin (or on the average firing rate over a particular time period) and therefore provides no information about the time course of neural tuning. In contrast, dPCA results in time-dependent components, which highlight the time course of neural tuning.
\item If a parameter has more than two possible values, such as the vibratory frequencies in the somatosensory working memory task, then reporting a percentage of neurons with significant tuning disregards the shape of the tuning curve. On the other hand, a vertical slice through any stimulus-dependent dPCA component results in a tuning curve. In the case of “rainbow-like” stimulus components (Figure 2B, 3B, or 5B), these tuning curves are easy to imagine.
\item To count significantly tuned neurons, one chooses an arbitrary p-value cutoff (e.g. $p<0.05$). This can potentially create a false impression that neurons are separated into two distinct sub-populations: those tuned to a given parameter (e.g. stimulus) and those that are not tuned. This, however, is not the case in any of the datasets considered here: each demixed component is expressed in the whole population, albeit with varying strength (Figure 6).
\item If neural tuning to several parameters is analyzed separately (e.g. with several t-tests or several one-way ANOVAs instead of one multi-way ANOVA), then the results of the tests can get confounded due to different number of trials in different conditions (“unbalanced” experimental design). In case of dPCA, all parameters are analyzed together, and the issue of confounding does not arise.
\end{enumerate}

An extended version of this approach uses multi-way ANOVA to test for firing rate modulation due to several parameters of interest, and repeats the test in a sliding time window to obtain the time course of neural tuning \citep[see e.g.][]{sul2010distinct, sul2011role}. This avoids the first and the fourth limitations listed above, but the other two limitations still remain.

\subsubsection*{II. Population averages}

In many studies, the time-varying firing rates, or PSTHs, of individual neurons are simply averaged over a subset of the neurons that were found to be significantly tuned to a particular parameter of interest \citep[e.g.][]{rainer1999prospective, kepecs2008neural}. While this approach can highlight some of the dynamics of the population response, it ignores the full complexity and heterogeneity of the data (Figure S1), which is largely averaged out. The approach thereby fails to faithfully represent the neural activities, and may lead to the false impression pointed out in the third issue above, namely that there are distinct subpopulations of neurons with distinct response patterns.

\subsubsection*{III. Demixing approach based on multiple regression}

\citet{mante2013context}, have recently introduced a demixing approach based on multiple regression. The authors first performed a linear regression of neural firing rates to several parameters of interest, and then took the vectors of regression coefficients as demixing axes. While this method proved sufficient for the purposes of that study, it does not aim to faithfully capture all of the data, which leads to several disadvantages. Specifically, the approach (a) ignores the condition-independent components, (b) assumes that all neural tuning is linear, (c) finds only a single demixed component for each parameter, and (d) cannot achieve demixing when the axes for different parameters are far from orthogonal. To illustrate these disadvantages, we applied this method to all our datasets and show the results in Figure S9.

A significant advantage of the approach by \citet{mante2013context}, is that it can deal with missing data or continuous task parameters, which dPCA currently can not. Future extensions of dPCA may therefore consider replacing the non-parametric dependencies of firing rates on task parameters with a parametric model, which would combine the advantages of both methods.

\subsubsection*{IV. Decoding approach}

Another multivariate approach relies on linear classifiers that predict the parameter of interest from the population firing rates. The strength of neural tuning can then be reported as the cross-validated classification accuracy. For example, \citet{meyers2012incorporation}, built linear classifiers for stimulus and match/non-match condition for the visuospatial working memory task analyzed above. Separate classifiers were used in each time bin, resulting in a time-dependent classification accuracy. The shape of the accuracy curve for the match/non-match condition follows the combined shapes of our decision components (and the same is true for stimulus). While the time-dependent classification accuracy provides an important and easily understandable summary of the population activity, it is far removed from the firing rates of individual neurons and is not directly representative of the neural tuning. The dPCA approach is more direct. 

\subsubsection*{V. Linear discriminant analysis}

Linear classifiers can also be used to inform the dimensionality reduction. For instance, linear discriminant analysis (LDA) reduces the dimensionality of a dataset while taking class labels into account. Whereas PCA looks for linear projections that maximize total variance (and ignores class labels), LDA looks for linear projections that maximize class separation, i.e. with maximal between-class and minimal within-class variance. Consequently, LDA is related to dPCA. However, LDA is a one-way technique (i.e. it considers only one parameter) and is not concerned with reconstruction of the original data, which makes it ill-suited for our purposes. See Supplementary Materials for an extended discussion on the differences between dPCA and LDA.

\subsection*{Insights obtained from applying dPCA to the four datasets}

One interesting outcome of the analysis concerns the strength of the condition-independent components. These components capture 70--90\% of the total, task-locked variance of neural firing rates. They are likely explained in part by an overall firing rate increase during certain task periods (e.g. during stimulus presentation). More speculatively, they could also be influenced by residual sensory or motor variables that vary rhythmically with the task, but are not controlled or monitored \citep{renart2014variability}. The attentional or motivational state of animals, for instance, often correlates with breathing \citep{huijbers2014respiration}, pupil dilation \citep{eldar2013effects}, body movements \citep{gouvea2014ongoing}, etc. 

The second important observation is that parameter tuning “moves” during the trial from one component to another. To visualize these movements, we set up  the dPCA algorithm such that it preferentially returns projections occupying only localized periods in time (see Methods). As a result, there are e.g. three separate stimulus components on Figure 2: one is active during the S1 period, one during the delay period, and one during the S2 period. The same can be observed in all other data sets as well: consider stimulus components in Figure 3 or decision components in Figures 4--5. The possibility to separate such components means that a neural subspace carrying information about a particular task parameter, changes (rotates) with time during the trial.

Our third finding is that most encoding axes turn out to be almost orthogonal to each other (only 22 out of 420 pairs are marked with stars on Figures 2--5E), even though dPCA, unlike PCA, does not enforce orthogonality. Non-orthogonality between two axes means that neurons expressing one of the components tend also to express the other one. Most examples of non-orthogonal pairs fall into the following three categories:

\begin{enumerate}
\item Non-orthogonality between a condition-dependent component and a condition-independent one (9 pairs), e.g. the first interaction (\#4) and the second condition-independent (\#2) components in Figure 4. This means that the neurons that are tuned to the presence/absence of reward will tend to have a specific time course of firing rate modulation, corresponding to component \#2. Note that the same effect is observed in Figure 5. 

\item Non-orthogonality between components describing one parameter (10 pairs), e.g. stimulus components in Figure 2. As many neurons express all three components, their axes end up being strongly non-orthogonal.

\item A special example is given by Figure 3, where the first stimulus component (\#7) and the first interaction component (\#10) are strongly non-orthogonal. As the interaction components in that particular data set are actually tuned to stimulus S2, this example can be seen a special case of the previous category.
\end{enumerate}

Only two pairs do not fall into any of the categories listed above (first stimulus and first decision components in both olfactory datasets). In all other cases, the condition-dependent components turn out to be almost orthogonal to each other. In other words, task parameters are represented independently of each other. Indeed, if latent components are independently, i.e. randomly mapped to a space of neurons, then the encoding axes will be nearly orthogonal to each other, because in a high-dimensional space any two random vectors are close to being orthogonal (unlike 2D or 3D cases). Substantial deviations from orthogonality can only occur if different parameters are encoded in a population in a non-independent way. Our results indicate that this is mostly not the case. In addition to that, orthogonal readouts are arguably the most effective, as they maximize signal-to-noise ratio, ensuring successful demixing in downstream brain areas.

\subsection*{Limitations}

One limitation of dPCA is that it works only with discrete parameters (and all possible parameter combinations must be present in the data) and would need to be extended to be able to treat continuous parameters as well. Another is that the number of neurons needs to be high enough: we found that at least $\sim$100 neurons are usually needed to achieve satisfactory demixing in the data considered here. Furthermore, here we worked with trial-averaged PSTHs and ignored trial-to-trial variability. The datasets presented here could not have been analyzed differently, because the neurons were recorded across multiple sessions, and noise correlations between neurons were therefore not known. Demixed PCA could in principle also be used for simultaneous recordings of sufficient dimensionality. However, dPCA does not specifically address the question of how to properly treat trial-to-trial variability, and this problem may require further extensions in the future.

\subsection*{Outlook}

Here we argued that dPCA is an exploratory data analysis technique that is well suited for neural data, as it provides a succinct and immediately accessible overview of the population activity. It is important to stress its exploratory nature: dPCA enables a researcher to have a look at the full data, ask further questions and perform follow-up analyses and statistical tests. It should therefore be a beginning, not the end of the analysis.

The dPCA code for Matlab and Python is available at \url{https://github.com/wielandbrendel/dPCA}.

\newpage
\section*{Materials and Methods}

\subsection*{Experimental data}
Brief descriptions of experimental paradigms are provided in the Results section and readers are referred to the original publications for all further details. Here we describe the selection of animals, sessions, and trials for the present manuscript. In all experiments neural recordings were obtained in multiple sessions, so most of the neurons were not recorded simultaneously.

\begin{enumerate}
\item Somatosensory working memory task in monkeys \citep{romo1999neuronal, brody2003timing}. We used the data from two monkeys (code names RR14 and RR15) that were trained with the same frequency set, and selected only the sessions where all six frequencies \{10, 14, 18, 26, 30, 34\} Hz were used for the first stimulation (other sessions were excluded). Monkeys made few mistakes (overall error rate was 6\%), and here we analyzed only correct trials. Monkey RR15 had additional 3 s delay after the end of the second stimulation before it was cued to provide the response. Using the data from monkey RR13 (that experienced different frequency set) led to very similar dPCA components (data not shown).

\item Visuospatial working memory task in monkeys \citep{qi2011changes, meyer2011stimulus, qi2012neural}. We used the data from two monkeys (code names AD and EL) that were trained with the same spatial task. Monkeys made few mistakes (overall error rate was 8\%), and here we analysed only correct trials. The first visual stimulus was presented at 9 possible spatial locations arranged in a 3$\times$3 grid (Figure 3A); here we excluded all the trials where the first stimulus was presented in the centre position.

\item Olfactory discrimination task in rats \citep{feierstein2006representation}. We used the data from all five rats (code names N1, P9, P5, T5, and W1). Some rats were trained with two distinct odours, some with four, some with six, and one rat experienced mixtures of two fixed odours in varying proportions. In all cases each odour was uniquely associated with one of the two available water ports (left/right). Following the original analysis \citep{feierstein2006representation}, we grouped all odours associated with the left/right reward together as a “left/right odour”. For most rats, caproic acid and 1-hexanol (shown on Figures 4--5A) were used as the left/right odour. We excluded from the analysis all trials that were aborted by rats before reward delivery (or before waiting 0.2 s at the reward port for the error trials).

\item Olfactory categorization task in rats \citep{kepecs2008neural}. We used the data from all three rats (code names N1, N48, and N49). Note that recordings from one of the rats (N1) were included in both this and previous datasets; when we excluded if from either of the datasets, the results stayed qualitatively the same (data not shown). We excluded from the analysis all trials that were aborted by rats before reward delivery (or before waiting 0.3 s at the reward port for the error trials).
\end{enumerate}

\subsection*{Neural recordings}
In each of the datasets each trial can be labeled with two parameters: “stimulus” and “decision”. Note that a “reward” label is not needed, because its value can be deduced from the other two due to the deterministic reward protocols in all tasks. For our analysis it is important to have recordings of each neuron in each possible condition (combination of parameters). Additionally, we required that in each condition there were at least $E_\mathrm{min} > 1$ trials, to reduce the standard error of the mean when averaging over trials, and also for cross-validation purposes. The cutoff was set to $E_\mathrm{min} = 5$ for both working memory datasets, and to $E_\mathrm{min} = 2$ for both olfactory datasets (due to less neurons available).

We have further excluded very few neurons with mean firing rate over 50 Hz, as neurons with very high firing rate can bias the variance-based analysis. Firing rates above 50 Hz were atypical in all datasets (number of excluded neurons for each dataset: 5 / 2 / 1 / 0). This exclusion had a minor positive effect on the components. We did not apply any variance-stabilizing transformations, but if the square-root transformation was applied, the results stayed qualitatively the same (data not shown).

No other pre-selection of neurons was used. This procedure left 832 neurons (230 / 602 for individual animals, order as above) in the somatosensory working memory dataset, 956 neurons (182 / 774) in the visuospatial working memory dataset, 437 neurons in the olfactory discrimination dataset (166 / 30 / 9 / 106 / 126), and 214 neurons in the olfactory categorization dataset (67 / 38 / 109).

\subsection*{Preprocessing}
Spike trains were filtered with a Gaussian kernel ($\sigma = 50$ ms) and averaged over all trials in each condition to obtain smoothed average peri-stimulus time histograms (PSTHs) for each neuron in each condition. 

In the visuospatial working memory dataset we identified the preferred direction of each neuron as the location that evoked maximum mean firing rate in the 500 ms time period while the first stimulus was shown. The directional tuning was shown before to have a symmetric bell shape \citep{qi2011changes, meyer2011stimulus}, with each neuron having its own preferred direction. We then re-sorted the trials (separately for each neuron) such that only 5 distinct stimuli were left: preferred direction, 45$^\circ$, 90$^\circ$, 135$^\circ$, and 180$^\circ$ away from the preferred direction.

In both olfactory datasets trials were self-paced. This means that trials could have very different length, making averaging of firing rates over trials impossible. We used the following re-stretching procedure (separately in each dataset) to equalize the length of all trials and to align several events of interest (see Figure S10 for the illustration). We defined five alignment events: odour poke in, odour poke out, water poke in, reward delivery, and water poke out. First, we aligned all trials on odour poke in ($T_1=0$) and computed median times of four other events $T_i,\, i=2\dots5$ (for the time of reward delivery we took the median over all correct trials). Second, we set $\Delta T$ to be the minimal waiting time between water port entry and reward delivery across the whole experiment ($\Delta T=0.2$ s for the olfactory discrimination task and $\Delta T=0.3$ s for the olfactory categorization task). Finally, for each trial we computed the PSTH $r(t)$ as described above, set $t_i,\, i=1\dots5$, to be the times of alignment events on this particular trial (for error trials we took $t_4 = t_3 + \Delta T$), and stretched $r(t)$ along the time axis in a piecewise-linear manner to align each $t_i$ with the corresponding $T_i$.

We made sure that this procedure did not introduce any artifacts by considering an alternative procedure, where short ($\pm$450ms) time intervals around each $t_i$ were cut out of each trial and concatenated together; this procedure is similar to the pooling of neural data performed in the original studies \citep{feierstein2006representation, kepecs2008neural}. The dPCA analysis revealed qualitatively similar components (Figure S6).

\subsection*{dPCA: marginalization}
For each neuron $n$ (out of $N$), stimulus $s$ (out of $S$), and decision $d$ (out of $Q$) we have a collection of $E \ge E_\mathrm{min}$ trials. For each trial we have a filtered spike train $r(t)$ sampled at $T$ time points. The number of trials $E$ can vary with $n$, $s$, and $d$, and so we denote it by $E_{nsd}$. For each neuron, stimulus, and decision we average over these $E_{nsd}$ trials to compute the mean firing rate $R_{nsd}(t)$. These data can be thought of as $C=SQ$ time-dependent neural trajectories (one for each condition) in a $N$-dimensional space $\mathbb R^N$ (Figure 1A). The number of distinct data points in this $N$-dimensional space is $SQT$. We collect them in one matrix $\X$ with dimensions $N \times SQT$ (i.e. $N$ rows, $SQT$ columns). 

Let $\X$ be centred, i.e. the mean response of any single neuron over all times and conditions is zero. As we previously showed in \citep{brendel2011demixed}, $\X$ can be decomposed into independent parts that we call “marginalizations” (Figure S11):
$$\X = \X_t + \X_{st} + \X_{dt} + \X_{sdt} = \sum \X_\phi.$$
Here $\X_t$ denotes the time-varying, but stimulus- and decision-invariant part of $\X$, which can be obtained by averaging $\X$ over all stimuli and decisions (i.e. over all conditions), i.e. $\X_t = \langle \X \rangle_{sd}$ (where the angle brackets denote the average over the subscripted parameters). The stimulus-dependent part $\X_{st} = \langle \X - \X_t \rangle_d$ is an average over all decisions of the part that is not explained by $\X_t$. Similarly, the decision-dependent part $\X_{dt} = \langle \X - \X_t \rangle_s$ is an average of the same remaining part over all stimuli. Finally, the higher-order, or “interaction”, term is calculated by subtracting all simpler terms, i.e. $\X_{sdt} = \X - \X_t - \X_{st} - \X_{dt}$. As a matrix, each marginalization has exactly the same dimensions as $\X$ (e.g. $\X_t$ is not a $N\times T$-dimensional matrix, but a $N\times SQT$ matrix with $SQ$ identical values for every time point).

In this manuscript we are not interested in separating neural activity that varies only with stimulus (but not with time) from neural activity that varies due to interaction of stimulus with time. More generally, however, we can treat all parameters on an equal footing. In this case, data labeled by $M$ parameters can be marginalized into $2^M$ marginalization terms. For $M = 3$ the most general decomposition is given by
$$\X = \X_0 + \X_s + \X_d + \X_t + \X_{sd} + \X_{dt} + \X_{st} + \X_{sdt}.$$
See Supplementary Materials for more details.

\subsection*{dPCA: separating time periods}
We have additionally separated stimulus, decision, and interaction components into those having variance in different time periods of the trial. Consider the somatosensory working memory task (Figure 2). Each trial can be reasonably split into three distinct periods: up until the end of S1 stimulation, delay period, and after beginning of S2 stimulation. We aimed at separating the components into those having variance in only one of these periods. For this, we further split e.g. $\X_{st}$ into $\X_{st}^{(1)}$, $\X_{st}^{(2)}$, $\X_{st}^{(3)}$ such that $\X_{st} = \X_{st}^{(1)} + \X_{st}^{(2)} + \X_{st}^{(3)}$ and each of the parts $\X_{st}^{(i)}$ equals zero outside of the corresponding time interval. This was done with stimulus, decision, and interaction marginalizations (but not with the condition-independent one).

As a result, three stimulus components shown on Figure 2B have very distinct time course. Note, however, that the angles between corresponding projection axes are far from orthogonal (Figure 2E). This means that most of the neurons expressing one of these components express other two as well, and so if the splitting had not been enforced, these components would largely have joined together (Figure S11).

For the visuospatial working memory dataset we split trials into four parts: before the end of S1, delay period, S2 period, second delay period. For both olfactory datasets we split trials into two parts on water poke in.

Importantly, we made sure that this procedure did not lead to any noticeable loss of explained variance, as compared to dPCA without time period separation (Figure S12). On the other hand, it often makes individual components easier to interpret (e.g. second stimulus components on Figures 2 and 3 are only active during the delay periods and so are clear “memory” components), and highlights the fact that parameter (e.g. stimulus) representations tend to rotate with time in the firing rate space.

\subsection*{dPCA: algorithm}

Given a decomposition $\X = \sum \X_\phi$, dPCA aims at finding directions in $\mathbb R^N$ that capture as much variance as possible, with an additional restriction that the variance in each direction should come from only one of $\X_\phi$. Let us assume that we want the algorithm to find $q$ directions for each marginalization (i.e. $4q$ directions in total). The cost function for dPCA is
$$L = \sum_\phi L_\phi = \sum_\phi \left(\|\X_\phi - \mathbf F_\phi \mathbf D_\phi \X\|^2 + \lambda\|\mathbf D_\phi\|^2 \right),$$
where $\mathbf F_\phi$ is an encoder matrix with $q$ columns and $\mathbf D_\phi$ is a decoder matrix with $q$ rows. Here and below matrix norm signifies Frobenius norm, i.e. $\|\X\|^2 = \sum_i \sum_j X_{ij}^2$. Without loss of generality we assume that $\mathbf F_\phi$ has orthonormal columns and that components are ordered such that their variance (row variance of $\mathbf D_\phi \X$) is decreasing. The term $\lambda \|\mathbf D_\phi\|$ regularizes the solution to avoid overfitting (Figure S13). 

For $\lambda = 0$ the optimization problem can be understood as a reduced rank regression problem \citep{izenman1975reduced, reinsel1998multivariate}: minimize $\|\X_\phi - \mathbf A_q \X\|$ with $\mathbf A_q = \mathbf F_\phi \mathbf D_\phi$ and $\mathrm{rank}(\mathbf A_q) \le q$. The minimum can be found in three steps:

\begin{enumerate}
\item Compute the standard regression solution $\mathbf A = \X_\phi \X^+$ to the unconstrained problem of minimizing $\|\X_\phi - \mathbf A\X\|$; 
\item Project $\mathbf A$ on the $q$-dimensional subspace $\mathbf U_q$ with highest variance to incorporate the rank constraint, i.e. $\mathbf A_q = \mathbf U_q \mathbf U_q^\top \mathbf A$, where $\mathbf U_q$ is a matrix of $q$ leading singular vectors of $\mathbf A \X$;
\item Factorize $\mathbf A_q = \mathbf F_\phi \mathbf D_\phi$ with $\mathbf F_\phi = \mathbf U_q$ and $\mathbf D_\phi = \mathbf U_q^\top \mathbf A$ to recover the encoder and decoder. 
\end{enumerate}
Finally, notice that the regularized problem $\lambda \ne 0$ can be reduced to the unregularized case by replacing $\mathbf X_\phi \to [\X_\phi \,|\, \mathbf 0]$ and $\X \to [\X \,|\, \lambda \mathbf I]$ where $\mathbf 0$ and $\mathbf I$ are $N\times N$ zero and unit matrices. See Supplementary Materials for a more detailed mathematical treatment.

We found that a very good approximation to the optimal solution can be achieved in a simpler way that can provide some further intuition. Let $\mathbf F_\phi$ be equal to a matrix of the first $J$ principal axes (singular vectors) of $\X_\phi$, join all $\mathbf F_\phi$ together horizontally to form one $N \times 4J$ matrix $\mathbf F$, and take $\mathbf D = \mathbf F^+$, pseudo-inverse of $\mathbf F$. This works well, provided that $J$ is chosen to capture most of the signal variance of $X_\phi$, but not larger. Choosing $J$ too small results in poor demixing, and choosing $J$ too large results in overfitting. We found that $J=10$ provides a good trade-off in all datasets considered here. However, the general method described above is a preferred approach, as it does not depend on the choice of $J$, provides a more accurate regularization and is derived from an explicit objective (hence avoids hidden assumptions).

This approximate solution highlights the conditions under which dPCA will work, i.e. will result in well-demixed components (components having variance in only one marginalization) that together capture most of the variance of the data. For this to work, the main principal axes of different marginalizations $\X_\phi$ should be non-collinear, or in other words, principal subspaces of different marginalizations should not overlap.

\subsection*{dPCA: regularization}
We used cross-validation to find the optimal regularization parameter $\lambda$  for each dataset. To separate the data into training and testing sets, we held out one random trial for each neuron in each condition as a set of $C=SQ$ test “pseudo-trials” $\X_\mathrm{test}$ (as the neurons were not recorded simultaneously, we do not have recordings of all $N$ neurons in any actual trial). Remaining trials were averaged to form a training set $\X_\mathrm{train}$. Note that $\X_\mathrm{test}$ and $\X_\mathrm{train}$ have the same dimensions. We then performed dPCA on $\X_\mathrm{train}$ for various values of $\lambda$, selected 10 components in each marginalization (i.e. 40 components in total) to obtain $\mathbf F_\phi(\lambda)$ and $\mathbf D_\phi(\lambda)$, and computed $$R(\lambda) = \frac{\sum_\phi\|\X_{\mathrm{train},\, \phi} - \mathbf F_\phi(\lambda) \mathbf D_\phi(\lambda) \X_\mathrm{test}\|^2 }{ \|\X_\mathrm{train}\|^2}$$ as a fraction of variance not explained by the held-out data. We repeated this procedure 10 times for different train-test splittings and averaged the resulting functions $R(\lambda)$. In all cases the average function $R(\lambda)$ had a clear minimum (Figure S14) that we selected as the optimal $\lambda$. An alternative formula for $R(\lambda)$ using only $\X_\mathrm{test}$ yielded the same results, see Supplementary Information. 

The values used for each dataset were $3.8\cdot 10^{-6}$ / $1.3\cdot 10^{-5}$ / $1.9\cdot 10^{-5}$ / $1.9\cdot 10^{-5}$ times the total variance $\|\X\|^2$ of the corresponding dataset.

\subsection*{dPCA: intuition on decoder and encoder}
It can be argued that only the decoding axes are of interest (and not the encoding axes). In the toy example shown on Figure 1F, decoding axes roughly correspond to discriminant axes that linear discriminant analysis (LDA) would find when trying to decode stimulus and time. This remains true in the case of real data as well (see Supplementary Materials). However, without encoding axes there is no way to reconstruct the original dataset and therefore no way in assigning “explained variance” to principal components.

On the other hand, it can be argued that only the encoding axes are of interest. Indeed, in the same toy example shown on Figure 1F, encoding axes roughly correspond to principal axes of the stimulus and time marginalizations. In other words, they show directions along which stimulus and time are varying the most. Correspondingly, instead of performing dPCA, one can perform standard PCA in each marginalization and analyze resulting components inside each marginalization (they will, by definition, be perfectly demixed). However, this makes a transformation data $\to$ dPC complex and involving a series of multi-trial averages. The brain has arguably to rely on standard linear projections to do any kind of trial-by-trial inference, and so to gain insight into the utility of the population code for a biological system, we prefer the method involving only linear projections of the full data. This is achieved with a decoder.

We believe therefore that both encoder and decoder are essential for our method and treat them on equal footing, in line with the schematic shown on Figure 1E.

\subsection*{Variance calculations}
The marginalization procedure ensures that the total $N\times N$ covariance matrix $\mathbf C$ is equal to a sum of covariance matrices from each marginalization:
$$\mathbf C = \mathbf C_{(t)} + \mathbf C_{(st)} + \mathbf C_{(dt)} + \mathbf C_{(sdt)}.$$
This means that the the variance of $\X$ in each direction $\mathbf d$ can be decomposed into a sum of variances due to to time, stimulus, decision, and stimulus-decision interaction. This fact was used to produce bar plots shown on Figures 2--5D. Namely, consider a dPC with a decoding vector $\mathbf d$ (that does not necessarily have a unit length). Total variance of this dPC is $S = \|\mathbf d^\top \X\|^2$ and it is equal to the sum of marginalized variances $S_\phi = \|\mathbf d^\top \X_\phi\|^2$.

This allows us to define a “demixing index” of each component as $\delta = \mathrm{max}\{S_\phi\}/S$. This index can range from 0.25 to 1, and the closer it is to 1, the better demixed the component is. As an example, for the somatosensory working memory dataset, the average demixing index over the first 15 PCA components is 0.76$\pm$0.16 (mean$\pm$SD), and over the first 15 dPCA components is 0.97$\pm$0.02, which means that dPCA achieves much better demixing ($p=0.00016$, Mann-Whitney-Wilcoxon ranksum test). For comparison, the average demixing index of individual neurons in this dataset is 0.55$\pm$0.18. In other datasets these numbers are similar.

To compute the cumulative variance explained by the first $q$ components, we cannot simply add up individual variances, as the demixing axes are not orthogonal. Instead, on Figures 2--5C we show “explained variance”, computed as follows: Let the decoding vectors for these axes be stacked as rows in a matrix $\mathbf D_q$ and encoding vectors as columns in a matrix $\mathbf F_q$. Then the proportion of total explained variance is given by
$$\frac{\|\X\|^2 - \|\X - \mathbf F_q \mathbf D_q \X\|^2 }{ \|\X\|^2}.$$

Note that for the standard PCA when $\mathbf F_q = \mathbf D_q^\top = \mathbf U_\mathrm{pca}$, the standard explained variance (sum of the first $q$ eigenvalues of the covariance matrix over the sum of all eigenvalues) can be given by an analogous formula: 
$$\frac{\|\X\|^2 - \|\X - \mathbf U_\mathrm{pca} \mathbf U_\mathrm{pca}^\top \X\|^2 }{ \|\X\|^2} = \frac{\|\mathbf U_\mathrm{pca}^\top \X\|^2 }{ \|\X\|^2}.$$

Following \citep{machens2010functional}, in Figures 2--5C we applied a correction to show the amount of explained “signal variance”. Assuming that each PSTH $r(t)$ consists of some trial-independent signal $s(t)$ and some random noise $\epsilon(t)$, $r(t) = s(t) + \epsilon(t)$, the average PSTH $R_{nsd}(t)$ will be equal to $R_{nsd}(t) = s(t) + E_{nsd}^{-1/2} \epsilon(t)$. If the number of trials $E_{nsd}$ is not very large, some of the resulting variance will be due to the noise term. To estimate this variance for each neuron in each condition, we took two random trials $r_{a1}(t)$ and $r_2(t)$ and considered $\Theta_{nsd}(t) = (2E_{nsd})^{-1/2}\big(r_1(t)-r_2(t)\big)$. These data form a data matrix $\boldsymbol \Theta$ of the same dimensions as $\X$, which has no signal but approximately the same amount of noise as $\X$. The following text assumes that $\boldsymbol \Theta$ is centred.

Singular values of $\boldsymbol \Theta$ provide an approximate upper bound of the amount of noise variance in each successive PCA or dPCA component. Therefore, the cumulative amount of signal variance for PCA is given by 
$$\frac{\sum_{i=1}^q \mu_i^2 - \sum_{i=1}^q \eta_i^2 }{ \|\X\|^2 - \|\boldsymbol \Theta\|^2},$$
where $\mu_i$ and $\eta_i$ are singular values of $\X$ and $\boldsymbol \Theta$ respectively. For dPCA, the formula becomes
$$\frac{\|\X\|^2 - \|\X - \mathbf F_q \mathbf D_q \X\|^2 - \sum_{i=1}^q \eta_i^2 }{ \|\X\|^2 - \|\boldsymbol \Theta\|^2}.$$

Pie charts in Figures 2--5D show the amount of signal variance in each marginalization. To compute it, we marginalize $\boldsymbol \Theta$, obtain a set of $\boldsymbol \Theta_\phi$, and then compute signal variance in marginalization $\phi$ as $\|\X_\phi\|^2 - \|\boldsymbol \Theta_\phi\|^2$. The sum over all marginalizations is equal to the total amount of signal variance $\|\X\|^2 - \|\boldsymbol \Theta\|^2$.

\subsection*{Angles between dPCs}
On Figures 2--5E stars mark the pairs of components whose encoding axes $\mathbf f_1$ and $\mathbf f_2$ are significantly and robustly non-orthogonal. These were identified as follows: In Euclidean space of $N$ dimensions, two random unit vectors (from a uniform distribution on the unit sphere $S^{N-1}$) have dot product (cosine of the angle between them) distributed with mean zero and standard deviation $N^{-1/2}$. For large $N$ the distribution is approximately Gaussian. Therefore, if $|\mathbf f_1 \cdot \mathbf f_2| > 3.3 \cdot N^{-1/2}$, we say that the axes are significantly non-orthogonal ($p<0.001$).

Coordinates of $\mathbf f_1$ quantify how much this component contributes to the activity of each neuron. Hence, if cells exhibiting one component also tend to exhibit another, the dot product between the axes $\mathbf f_1 \cdot \mathbf f_2 > 0$ is positive (note that $\mathbf f_1 \cdot \mathbf f_2$ is approximately equal to the correlation between the coordinates of $\mathbf f_1$ and $\mathbf f_2$). Sometimes, however, the dot product has large absolute value only due to several outlying cells. To ease interpretation, we marked with stars only those pairs of axes for which the absolute value of Spearman (robust) correlation was above 0.2 with $p<0.001$ (in addition to the above criterion on $\mathbf f_1 \cdot \mathbf f_2$).

\subsection*{Decoding Accuracy and Cross-Validation}
We used decoding axis $\mathbf d$ of each dPC in stimulus, decision, and interaction marginalizations as a linear classifier to decode stimulus, decision, or condition respectively. Black lines on Figures 2--5B show time periods of significant classification. A more detailed description follows below.

We used 100 iterations of stratified Monte Carlo leave-group-out cross-validation, where on each iteration we held out one trial for each neuron in each condition as a set of $C=SQ$ test “pseudo-trials” $\X_\mathrm{test}$ and averaged over remaining trials to form a training set $\X_\mathrm{train}$ (see above). After running dPCA on $\X_\mathrm{train}$, we used decoding axes of the first three stimulus/decision/interaction dPCs as a linear classifier to decode stimulus/decision/condition respectively. Consider e.g. the first stimulus dPC: first, for each stimulus, we computed the mean value of this dPC separately for every time-point. Then we projected each test trial on the corresponding decoding axis and classified it at each time-point according to the closest class mean. Proportion of test trials (out of $C$) classified correctly resulted in a time-dependent classification accuracy, which we averaged over 100 cross-validation iterations. Note that this is a stratified procedure: even though in reality some conditions have much fewer trials than others, here we classify exactly the same number of “pseudo-trials” per condition. At the same time, as the coordinates of individual data points in each pseudo-trial are pooled from different sessions, the influence of noise correlations on the classification accuracies is neglected, similar to \citet{meyers2012incorporation}.

We then used 100 shuffles to compute Monte Carlo distribution of classification accuracies expected by chance. On each iteration for each neuron we shuffled all available trials between conditions, respecting the number of trials per condition (i.e. all $\sum_{sd} E_{nsd}$ trials were shuffled and then randomly assigned to the conditions such that all values $E_{nsd}$ stayed the same). Then exactly the same classification procedure as above (with 100 cross-validation iterations) was applied to the shuffled dataset to find mean classification accuracy for the first stimulus, decision, and interaction components. All 100 shuffling iterations resulted in a set of 100 time-dependent accuracies expected by chance.

The time periods when actual classification accuracy exceeded all 100 shuffled decoding accuracies in at least 10 consecutive time bins are marked by black lines on Figures 2--5. Components without any periods of significant classification are not shown. See Figure S4 for classification accuracies in each dataset. Monte Carlo computations took $\sim$24 hours for each of the larger datasets on a 6 core 3.2 Ghz Intel i7-3930K processor.

\newpage
\bibliography{main}

\begin{thebibliography}{}

\bibitem[Barak et~al., 2010]{barak2010neuronal}
Barak, O., Tsodyks, M., and Romo, R. (2010).
\newblock Neuronal population coding of parametric working memory.
\newblock {\em The Journal of Neuroscience}, 30(28):9424--9430.

\bibitem[Brendel et~al., 2011]{brendel2011demixed}
Brendel, W., Romo, R., and Machens, C.~K. (2011).
\newblock Demixed principal component analysis.
\newblock In {\em Advances in Neural Information Processing Systems}, pages
  2654--2662.

\bibitem[Brody et~al., 2003]{brody2003timing}
Brody, C.~D., Hern{\'a}ndez, A., Zainos, A., and Romo, R. (2003).
\newblock Timing and neural encoding of somatosensory parametric working memory
  in macaque prefrontal cortex.
\newblock {\em Cerebral Cortex}, 13(11):1196--1207.

\bibitem[Buesing et~al., 2012]{buesing2012spectral}
Buesing, L., Sahani, M., and Macke, J.~H. (2012).
\newblock Spectral learning of linear dynamics from generalised-linear
  observations with application to neural population data.
\newblock In {\em Advances in neural information processing systems}, pages
  1682--1690.

\bibitem[Churchland et~al., 2012]{churchland2012neural}
Churchland, M., Cunningham, J., Kaufman, M., Foster, J., Nuyujukian, P., Ryu,
  S., and Shenoy, K. (2012).
\newblock Neural population dynamics during reaching.
\newblock {\em Nature}, 487(7405):51--56.

\bibitem[Cunningham and Yu, 2014]{cunningham2014dimensionality}
Cunningham, J.~P. and Yu, B.~M. (2014).
\newblock Dimensionality reduction for large-scale neural recordings.
\newblock {\em Nature Neuroscience}.

\bibitem[Durstewitz, 2004]{durstewitz2004neural}
Durstewitz, D. (2004).
\newblock Neural representation of interval time.
\newblock {\em Neuroreport}, 15(5):745.

\bibitem[Eldar et~al., 2013]{eldar2013effects}
Eldar, E., Cohen, J.~D., and Niv, Y. (2013).
\newblock The effects of neural gain on attention and learning.
\newblock {\em Nature Neuroscience}, 16(8):1146--1153.

\bibitem[Feierstein et~al., 2006]{feierstein2006representation}
Feierstein, C.~E., Quirk, M.~C., Uchida, N., Sosulski, D.~L., and Mainen, Z.~F.
  (2006).
\newblock Representation of spatial goals in rat orbitofrontal cortex.
\newblock {\em Neuron}, 51(4):495--507.

\bibitem[Gouv{\^e}a et~al., 2014]{gouvea2014ongoing}
Gouv{\^e}a, T.~S., Monteiro, T., Soares, S., Atallah, B.~V., and Paton, J.~J.
  (2014).
\newblock Ongoing behavior predicts perceptual report of interval duration.
\newblock {\em Frontiers in Neurorobotics}, 8.

\bibitem[Hastie et~al., 2009]{hastie2009elements}
Hastie, T., Tibshirani, R., Friedman, J., Hastie, T., Friedman, J., and
  Tibshirani, R. (2009).
\newblock {\em The elements of statistical learning}.
\newblock Springer.

\bibitem[Hern{\'a}ndez et~al., 2010]{hernandez2010decoding}
Hern{\'a}ndez, A., N{\'a}cher, V., Luna, R., Zainos, A., Lemus, L., Alvarez,
  M., V{\'a}zquez, Y., Camarillo, L., and Romo, R. (2010).
\newblock Decoding a perceptual decision process across cortex.
\newblock {\em Neuron}, 66(2):300--314.

\bibitem[Huijbers et~al., 2014]{huijbers2014respiration}
Huijbers, W., Pennartz, C., Beldzik, E., Domagalik, A., Vinck, M., Hofman,
  W.~F., Cabeza, R., and Daselaar, S.~M. (2014).
\newblock Respiration phase-locks to fast stimulus presentations: Implications
  for the interpretation of posterior midline “deactivations”.
\newblock {\em Human Brain Mapping}, 35(9):4932--4943.

\bibitem[Izenman, 1975]{izenman1975reduced}
Izenman, A.~J. (1975).
\newblock Reduced-rank regression for the multivariate linear model.
\newblock {\em Journal of Multivariate Analysis}, 5(2):248--264.

\bibitem[Janssen and Shadlen, 2005]{janssen2005representation}
Janssen, P. and Shadlen, M.~N. (2005).
\newblock A representation of the hazard rate of elapsed time in macaque area
  lip.
\newblock {\em Nature Neuroscience}, 8(2):234--241.

\bibitem[Jun et~al., 2010]{jun2010heterogenous}
Jun, J.~K., Miller, P., Hern{\'a}ndez, A., Zainos, A., Lemus, L., Brody, C.~D.,
  and Romo, R. (2010).
\newblock Heterogenous population coding of a short-term memory and decision
  task.
\newblock {\em The Journal of Neuroscience}, 30(3):916--929.

\bibitem[Kepecs et~al., 2008]{kepecs2008neural}
Kepecs, A., Uchida, N., Zariwala, H.~A., and Mainen, Z.~F. (2008).
\newblock Neural correlates, computation and behavioural impact of decision
  confidence.
\newblock {\em Nature}, 455(7210):227--231.

\bibitem[Komura et~al., 2001]{komura2001retrospective}
Komura, Y., Tamura, R., Uwano, T., Nishijo, H., Kaga, K., and Ono, T. (2001).
\newblock Retrospective and prospective coding for predicted reward in the
  sensory thalamus.
\newblock {\em Nature}, 412(6846):546--549.

\bibitem[Machens, 2010]{machens2010demixing}
Machens, C.~K. (2010).
\newblock Demixing population activity in higher cortical areas.
\newblock {\em Frontiers in Computational Neuroscience}, 4.

\bibitem[Machens et~al., 2005]{machens2005flexible}
Machens, C.~K., Romo, R., and Brody, C.~D. (2005).
\newblock Flexible control of mutual inhibition: a neural model of two-interval
  discrimination.
\newblock {\em Science}, 307(5712):1121--1124.

\bibitem[Machens et~al., 2010]{machens2010functional}
Machens, C.~K., Romo, R., and Brody, C.~D. (2010).
\newblock Functional, but not anatomical, separation of “what” and
  “when” in prefrontal cortex.
\newblock {\em The Journal of Neuroscience}, 30(1):350--360.

\bibitem[Mante et~al., 2013]{mante2013context}
Mante, V., Sussillo, D., Shenoy, K.~V., and Newsome, W.~T. (2013).
\newblock Context-dependent computation by recurrent dynamics in prefrontal
  cortex.
\newblock {\em Nature}, 503(7474):78--84.

\bibitem[Meyer et~al., 2011]{meyer2011stimulus}
Meyer, T., Qi, X.-L., Stanford, T.~R., and Constantinidis, C. (2011).
\newblock Stimulus selectivity in dorsal and ventral prefrontal cortex after
  training in working memory tasks.
\newblock {\em The Journal of Neuroscience}, 31(17):6266--6276.

\bibitem[Meyers et~al., 2012]{meyers2012incorporation}
Meyers, E.~M., Qi, X.-L., and Constantinidis, C. (2012).
\newblock Incorporation of new information into prefrontal cortical activity
  after learning working memory tasks.
\newblock {\em Proceedings of the National Academy of Sciences},
  109(12):4651--4656.

\bibitem[Pagan and Rust, 2014]{pagan2014quantifying}
Pagan, M. and Rust, N.~C. (2014).
\newblock Quantifying the signals contained in heterogeneous neural responses
  and determining their relationships with task performance.
\newblock {\em Journal of Neurophysiology}, 112(6):1584--1598.

\bibitem[Park et~al., 2014]{park2014encoding}
Park, I.~M., Meister, M.~L., Huk, A.~C., and Pillow, J.~W. (2014).
\newblock Encoding and decoding in parietal cortex during sensorimotor
  decision-making.
\newblock {\em Nature Neuroscience}, 17:1395--1403.

\bibitem[Pfau et~al., 2013]{pfau2013robust}
Pfau, D., Pnevmatikakis, E.~A., and Paninski, L. (2013).
\newblock Robust learning of low-dimensional dynamics from large neural
  ensembles.
\newblock In {\em Advances in Neural Information Processing Systems}, pages
  2391--2399.

\bibitem[Qi et~al., 2011]{qi2011changes}
Qi, X.-L., Meyer, T., Stanford, T.~R., and Constantinidis, C. (2011).
\newblock Changes in prefrontal neuronal activity after learning to perform a
  spatial working memory task.
\newblock {\em Cerebral Cortex}, 21(12):2722--2732.

\bibitem[Qi et~al., 2012]{qi2012neural}
Qi, X.-L., Meyer, T., Stanford, T.~R., and Constantinidis, C. (2012).
\newblock Neural correlates of a decision variable before learning to perform a
  match/non-match task.
\newblock {\em The Journal of Neuroscience}, 32(18):6161--6169.

\bibitem[Rainer et~al., 1999]{rainer1999prospective}
Rainer, G., Rao, S.~C., and Miller, E.~K. (1999).
\newblock Prospective coding for objects in primate prefrontal cortex.
\newblock {\em The Journal of Neuroscience}, 19(13):5493--5505.

\bibitem[Reinsel and Velu, 1998]{reinsel1998multivariate}
Reinsel, G.~C. and Velu, R.~P. (1998).
\newblock {\em Multivariate reduced-rank regression}.
\newblock Springer.

\bibitem[Renart and Machens, 2014]{renart2014variability}
Renart, A. and Machens, C.~K. (2014).
\newblock Variability in neural activity and behavior.
\newblock {\em Current Opinion in Neurobiology}, 25:211--220.

\bibitem[Rigotti et~al., 2013]{rigotti2013importance}
Rigotti, M., Barak, O., Warden, M.~R., Wang, X.-J., Daw, N.~D., Miller, E.~K.,
  and Fusi, S. (2013).
\newblock The importance of mixed selectivity in complex cognitive tasks.
\newblock {\em Nature}, 497(7451):585--590.

\bibitem[Rodriguez and Laio, 2014]{rodriguez2014clustering}
Rodriguez, A. and Laio, A. (2014).
\newblock Clustering by fast search and find of density peaks.
\newblock {\em Science}, 344(6191):1492--1496.

\bibitem[Romo et~al., 1999]{romo1999neuronal}
Romo, R., Brody, C.~D., Hern{\'a}ndez, A., and Lemus, L. (1999).
\newblock Neuronal correlates of parametric working memory in the prefrontal
  cortex.
\newblock {\em Nature}, 399(6735):470--473.

\bibitem[Romo and Salinas, 2003]{romo2003flutter}
Romo, R. and Salinas, E. (2003).
\newblock Flutter discrimination: neural codes, perception, memory and decision
  making.
\newblock {\em Nature Reviews Neuroscience}, 4(3):203--218.

\bibitem[Singh and Eliasmith, 2006]{singh2006higher}
Singh, R. and Eliasmith, C. (2006).
\newblock Higher-dimensional neurons explain the tuning and dynamics of working
  memory cells.
\newblock {\em The Journal of Neuroscience}, 26(14):3667--3678.

\bibitem[Sul et~al., 2011]{sul2011role}
Sul, J.~H., Jo, S., Lee, D., and Jung, M.~W. (2011).
\newblock Role of rodent secondary motor cortex in value-based action
  selection.
\newblock {\em Nature Neuroscience}, 14(9):1202--1208.

\bibitem[Sul et~al., 2010]{sul2010distinct}
Sul, J.~H., Kim, H., Huh, N., Lee, D., and Jung, M.~W. (2010).
\newblock Distinct roles of rodent orbitofrontal and medial prefrontal cortex
  in decision making.
\newblock {\em Neuron}, 66(3):449--460.

\bibitem[Uchida and Mainen, 2003]{uchida2003speed}
Uchida, N. and Mainen, Z.~F. (2003).
\newblock Speed and accuracy of olfactory discrimination in the rat.
\newblock {\em Nature Neuroscience}, 6(11):1224--1229.

\bibitem[Wang et~al., 2013]{wang2013dorsomedial}
Wang, A.~Y., Miura, K., and Uchida, N. (2013).
\newblock The dorsomedial striatum encodes net expected return, critical for
  energizing performance vigor.
\newblock {\em Nature Neuroscience}, 16(5):639--647.

\bibitem[Wohrer et~al., 2013]{wohrer2013population}
Wohrer, A., Humphries, M.~D., and Machens, C.~K. (2013).
\newblock Population-wide distributions of neural activity during perceptual
  decision-making.
\newblock {\em Progress in Neurobiology}, 103:156--193.

\bibitem[Yu et~al., 2009]{yu2009gaussian}
Yu, B., Cunningham, J., Santhanam, G., Ryu, S., Shenoy, K., and Sahani, M.
  (2009).
\newblock Gaussian-process factor analysis for low-dimensional single-trial
  analysis of neural population activity.
\newblock {\em Journal of Neurophysiology}, 102(1):614--635.

\end{thebibliography}



\section*{Supplementary material}

# Demixed principal component analysis of population activity in higher cortical areas reveals independent representation of task parameters

## Supplementary Information

D Kobak*, W Brendel*, C Constantinidis, C Feierstein, A Kepecs, Z Mainen,
R Romo, X-L Qi, N Uchida, C Machens

# List of Figures

# Contents

# S1 Supplementary Figures

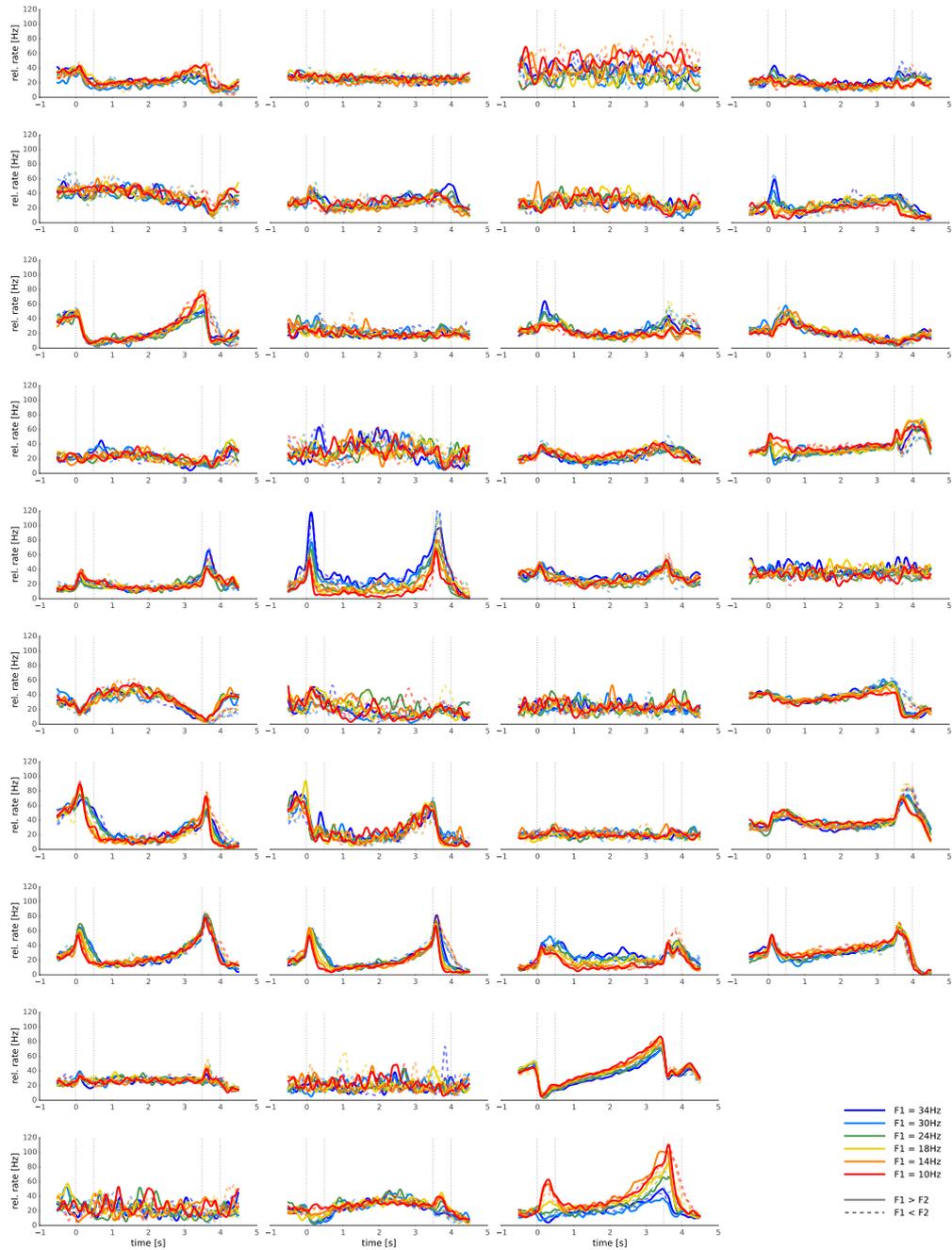

**Figure S1. Heterogeneity of single neuron responses.** Fourty randomly chosen neurons from the somatosensory working memory task in monkeys with average firing rates between 20 Hz and 40 Hz. Note large number of neurons showing mixed selectivity.

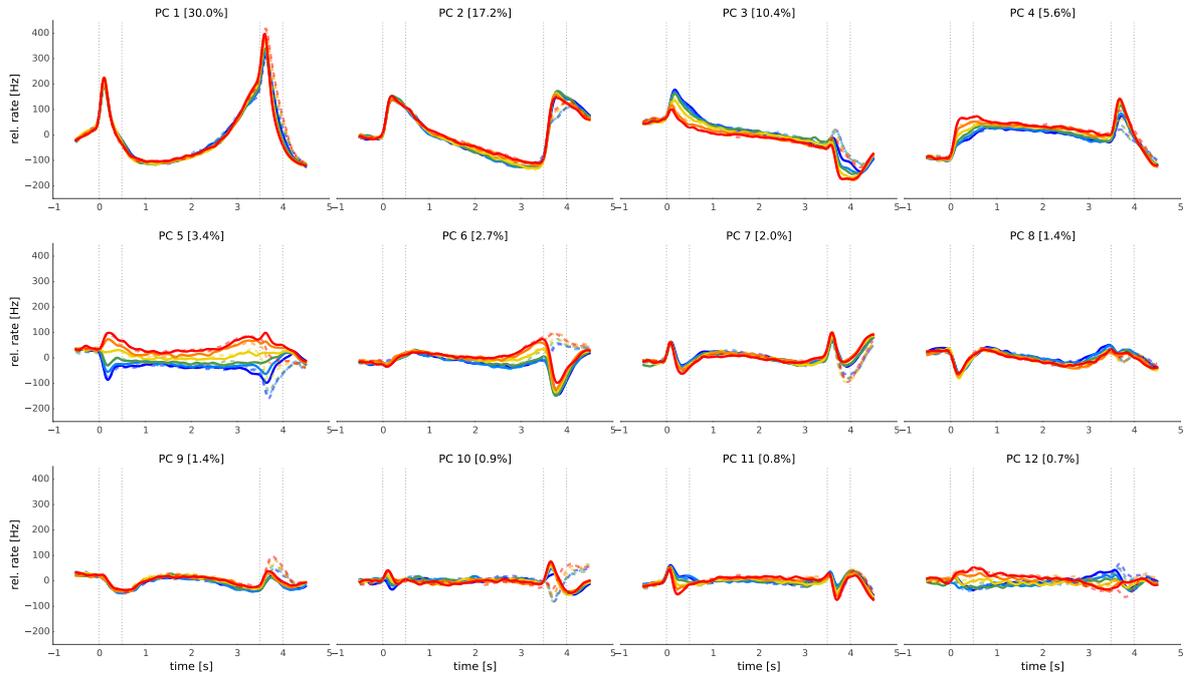

**Figure S2. Standard PCA of neural data.** First twelve principal components (standard PCA) in the somatosensory working memory task in monkeys. Legend equivalent to Figure S1. Note that some components strongly exhibit mixed selectivity. To quantify this effect, we can consider a "demixing index" of each component (see Methods for definition). For the first 15 PCA components, the average demixing index is $0.76 \pm 0.16$ (mean±SD). For comparison, for the first 15 dPCA components (Figure 2) the average demixing ratio is $0.97 \pm 0.02$. The difference is highly significant ($p = 0.00016$, Mann-Whitney-Wilcoxon ranksum test).

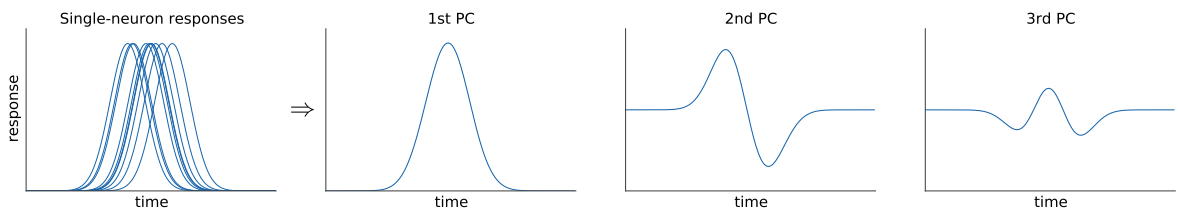

**Figure S3. Fourier-like artifacts in PCA.** (Left) In this toy example, single neuron responses are generated from the same underlying Gaussian but are randomly shifted in time. Such a situation can easily occur in real data either because different neurons have slightly different response latencies, or because they were recorded in different sessions. (Right) First three PCA components of the population data. While the leading component resembles the true signal, higher order components look like higher Fourier harmonics. They are artifacts of the jitter in time. See also section S2.5.

3

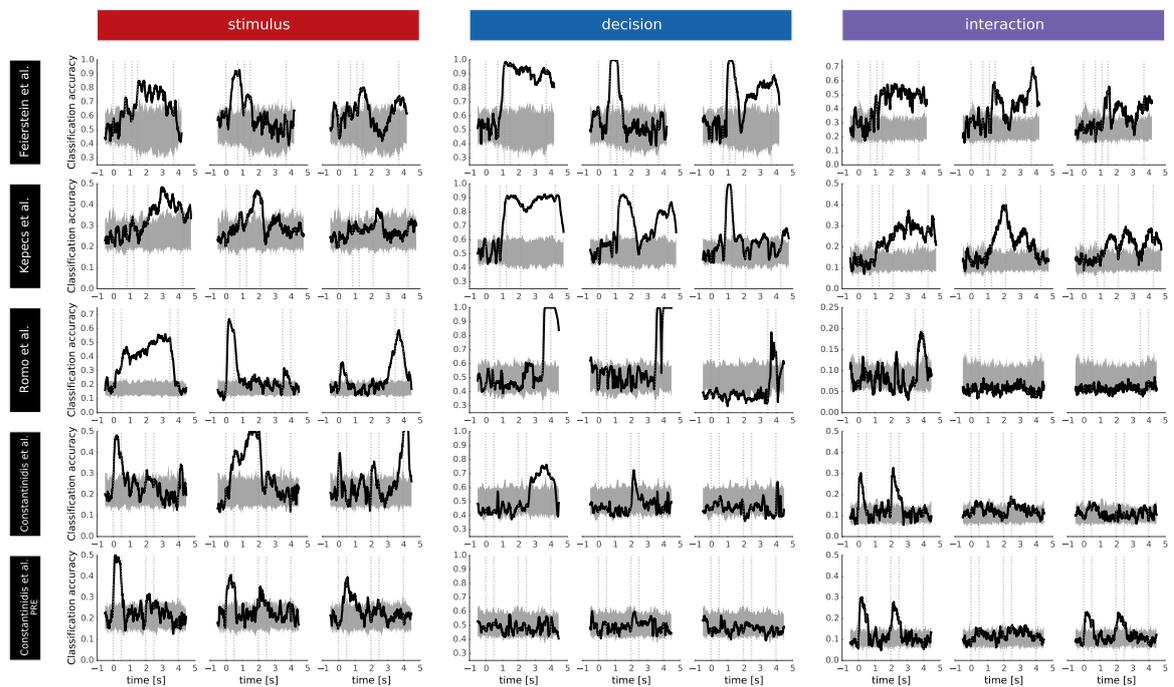

**Figure S4. Classification accuracies.** For each data set (rows) we performed cross-validation to quantify the classification accuracy of a linear classifier (black lines) given by the first three stimulus/decision/interaction dPCs (columns). Shaded gray regions show distribution of classification accuracies expected by chance (as estimated by 100 iterations of shuffling procedure). See Materials and Methods in the main text for the details.

4

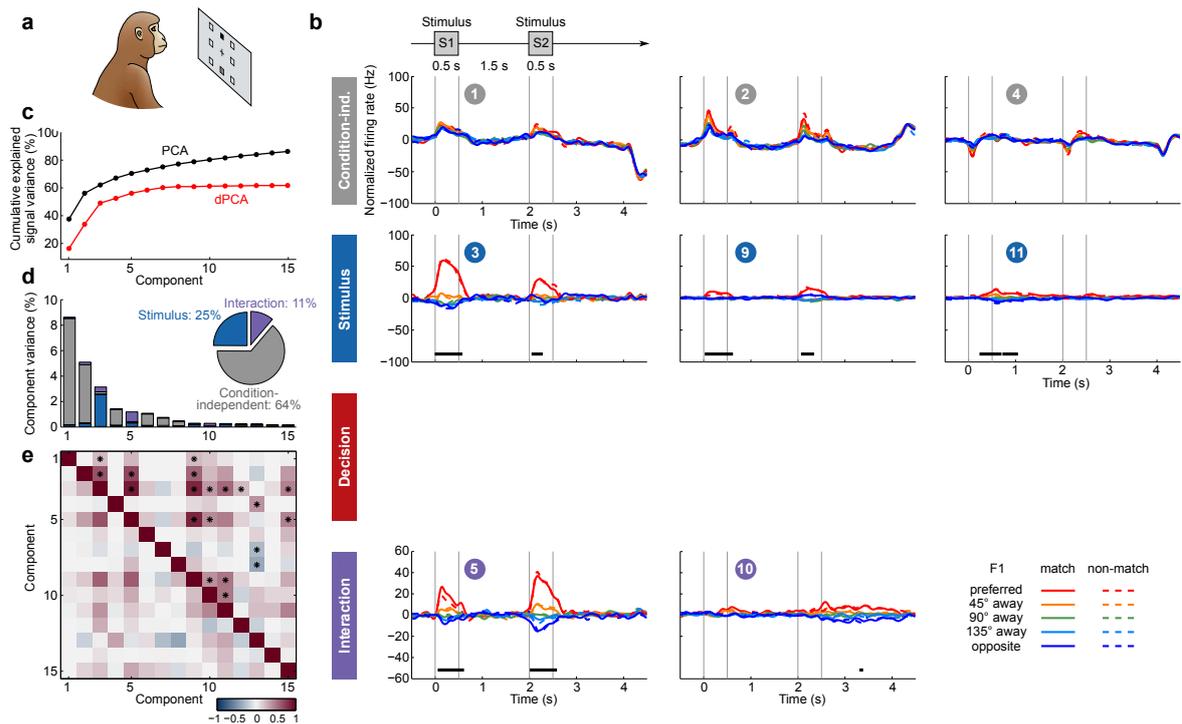

**Figure S5. Visuospatial working memory task in monkeys before training.** Demixed PCA analysis of the visuospatial working memory task in monkeys before training. Compare with Figure 3 of the main text and see the caption there. This analysis was done with 673 neurons from the same two monkeys as in the main text (number of neurons per monkey: 105 / 568; one neuron was excluded due to mean firing rate exceeding 50 Hz, see Methods). Note that many components come out very much non-orthogonal, with angles between them as small as 27 degrees: this is true for stimulus and interaction components (note that interaction components are tuned to the S2 stimulus, see main text), and also for stimulus components themselves (due to time period separating, see Methods). We could get clearer results, with much more orthogonal and also much more de-correlated components, if we do not enforce time period separation and do not demix stimulus from stimulus-decision interaction. In this case the leading dPCA component is the first stimulus component with a pronounced "sensory" activity in both S1 and S2 periods. However, here we show the analysis that can be directly compared to Figure 3.

5

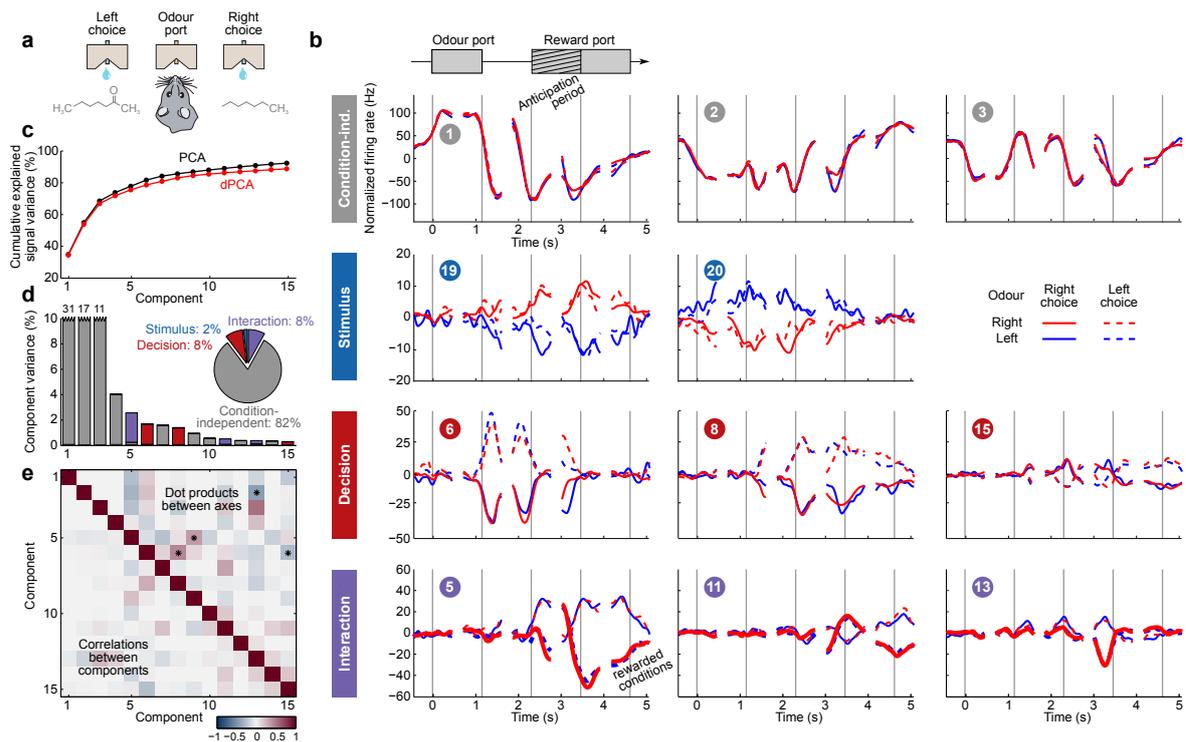

**Figure S6. dPCA of olfactory discrimination task in rats without re-stretching** Here, using data from the olfactory discrimination task in rats, we demonstrate that applying dPCA to the non-re-stretched data leads to qualitatively the same results as applying it to the re-stretched data as was done in the main text, see Figure 4. The purpose of re-stretching was to average over trials while aligning on five events of interest: odour port enter, odour port exit, reward port enter, reward delivery, and reward port exit (see Methods). Here we average over trials by cutting out a ±450 ms time interval around each event of interest and concatenating these cuts together (so that each trial results in a $450 \cdot 2 \cdot 5 = 4500$ ms long "piecewise" PSTH). Then we apply the same dPCA procedure (the optimal value of regularization parameter $\lambda$ for these data, when normalized by the total variance, was $1.3 \cdot 10^{-5}$, close to that for the re-stretched data). See Figure 4 in the main text for the full caption. Vertical gray lines on all subplots in (b) show alignment events, as on Figure 4. Dashed black lines show time points of cuts, e.g. mark discontinueties in time.

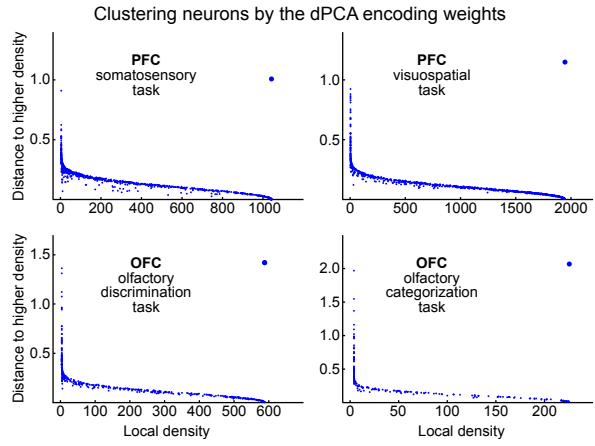

**Figure S7. Clustering of encoding weights by density peaks.** Each subplot shows one data set. In each case we performed dPCA, took first 15 components, and ran the recently developed clustering algorithm from (Rodriguez and Laio, 2014) to cluster the neurons in the 15-dimensional space of with coordinates given by the rows of encoder matrix **F**. The clustering algorithm works in two steps: first, we compute local density for each point (i.e. for each neuron), using Gaussian kernel with $\sigma = 0.1$. Second, for each point we find the minimal distance to a point with higher local density (if there is no such point, then take the distance to the furthest point). Each subplot shows local density plotted against distance to the next point with higher density for each of the $N$ neurons. Cluster centres should be outliers in the upper-right corner. In each case, there is only one such outlier, indicating a single cluster.

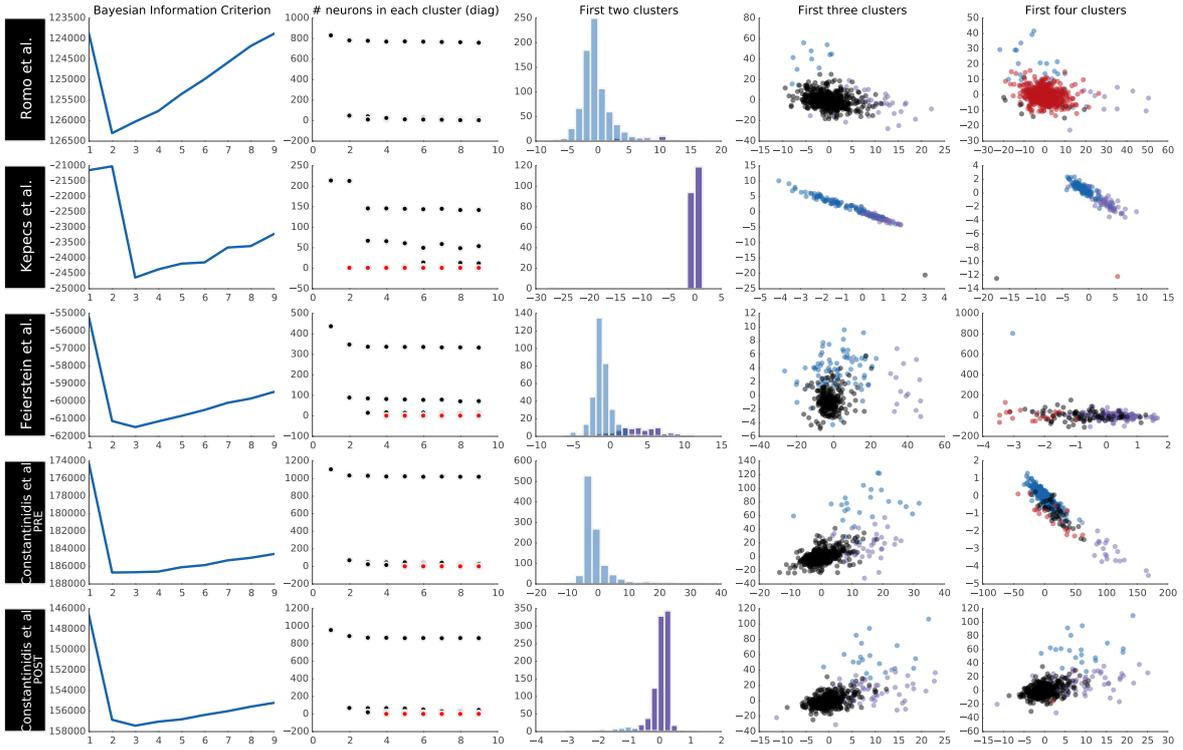

**Figure S8. Clustering of encoding weights with Gaussian Mixture Models.** Each row shows one data set. In each case we performed dPCA, took 10 components in each marginalization (40 columns in total), and attempted to cluster $N$ points (one for each neuron) in the 40-dimensional space with coordinates given by the rows of encoder matrix $\mathbf{F}$. We fitted Gaussian Mixture Model (GMM) with $K$ Gaussians and diagonal covariance matrices $\Sigma_k$. First column: Bayesian Information Criterion (BIC) as a function of the number of clusters. In each case BIC indicated that 2–3 clusters provide the optimal fit to the data. We argue, however, that this outcome is misleading in the sense that there are no separate clusters at all (so the optimal $K$ is in fact equal to 1), see below. Second column: the number of neurons per cluster (red dots denote clusters with only a single neuron). Third column: we fitted GMM with two clusters, projected the data on the subspace with strongest separation of two clusters (obtained using linear discriminant analysis, LDA), and show distributions of values in each cluster. Note that "clusters" are hardly separated. Fourth and fifth column: equivalent to the third column but for GMM fitted with three and four clusters respectively, and data projected onto two-dimensional LDA subspace. Note again that the clusters are hardly separated; instead there is one cloud of points that is not perfectly Gaussian and so GMM chooses to fit it with several "clusters".

8

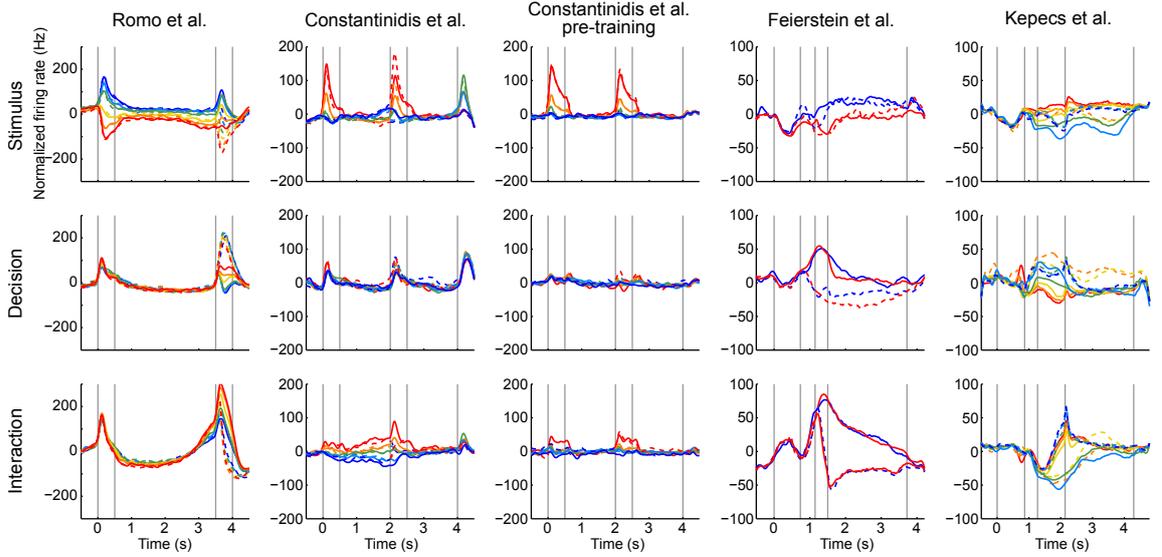

**Figure S9. Demixing approach of Mante et al.** We applied the demixing approach from (Mante et al., 2013) to all data sets considered in this manuscript. Briefly, the algorithm of Mante et al. is as follows. (1) Perform PCA of the trial-average neural data (as in our manuscript, see Figure S2) and define "denoising" matrix $\mathbf{K}$ as a linear projector on the space spanned by the first 12 principal axes: $\mathbf{K} = \mathbf{U}_{12}\mathbf{U}_{12}^\top$. (2) For each neuron $i$, regress its firing rate at each time point $t$ on stimulus, decision, and interaction between them: $x_i(t) = \beta_i^o(t) + \beta_i^s(t)s + \beta_i^d(t)d + \beta_i^{sd}(t)sd + \epsilon$, where $d$ is any suitable parametrization of decision, e.g. $d \in [-1, +1]$, and $s$ is stimulus value (e.g. actual stimulation frequency in the somatosensory working memory task in monkeys). (3) Take $N$ values of $\beta_i^s(t)$ as defining a vector $\boldsymbol{\beta}^s(t) \in \mathbb{R}^N$, and analogously define $\boldsymbol{\beta}^d(t)$ and $\boldsymbol{\beta}^{sd}(t)$. (4) Project each $\boldsymbol{\beta}^\phi(t)$ onto the space spanned by the first 12 PCA axes: $\boldsymbol{\beta}^\phi(t) \to \mathbf{K}\boldsymbol{\beta}^\phi(t)$. (5) For each of the three parameters select one vector $\boldsymbol{\beta}_*^\phi = \boldsymbol{\beta}^\phi(t_{\max}^\phi)$ where $t_{\max}^\phi$ is the time point at which the norm of $\boldsymbol{\beta}^\phi(t)$ is maximal. (6) Finally, stack three obtained vectors to form a matrix $\mathbf{B} = [\boldsymbol{\beta}_*^s, \boldsymbol{\beta}_*^d, \boldsymbol{\beta}_*^{sd}]$ and perform a QR decomposition $\mathbf{B} = \mathbf{QR}$ with $\mathbf{Q}$ orthonormal and $\mathbf{R}$ upper triangular to find a set of three orthogonal demixing axes. If $\boldsymbol{\beta}_*^s$, $\boldsymbol{\beta}_*^d$, and $\boldsymbol{\beta}_*^{sd}$ are far from orthogonal, then the resulting axes will strongly depend on the order of stacking them into the matrix $\mathbf{B}$. We found that we obtain best results if we select this order *ad hoc* for each data set; the orders we used to produce this figure were stimulus→decision→interaction for the somatosensory working memory task, stimulus→interaction→decision for the visuospatial working memory task, and interaction→decision→stimulus for both olfactory tasks.

9

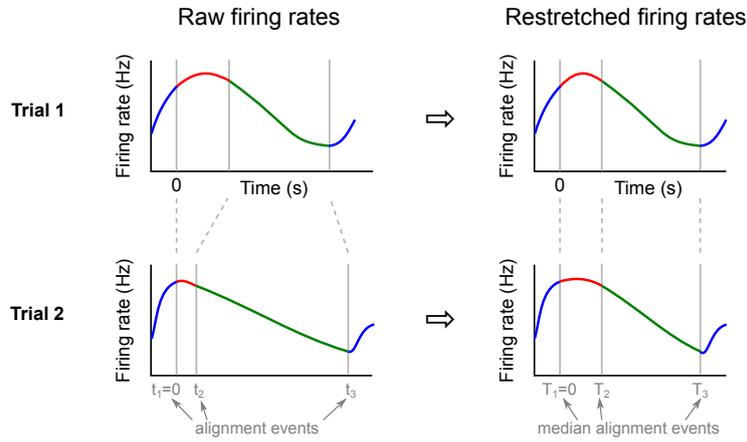

**Figure S10. Restretching procedure.** When analyzing the self-paced olfactory datasets, we used the following re-stretching procedure to equalize the length of all trials and to align several events of interest. We defined several alignment events (such as odour poke in, odour poke out, etc.) and for each trial found the times $t_i$ of these events. After aligning all trials on $t_1 = 0$ (left) we computed median times $T_i$ for all other events. Then for each trial the firing rate we re-stretched on each interval $[t_i, t_{i+1}]$ to align it with $[T_i, T_{i+1}]$ (right). After such restretching, all events are aligned and the trials corresponding to one condition can be averaged.

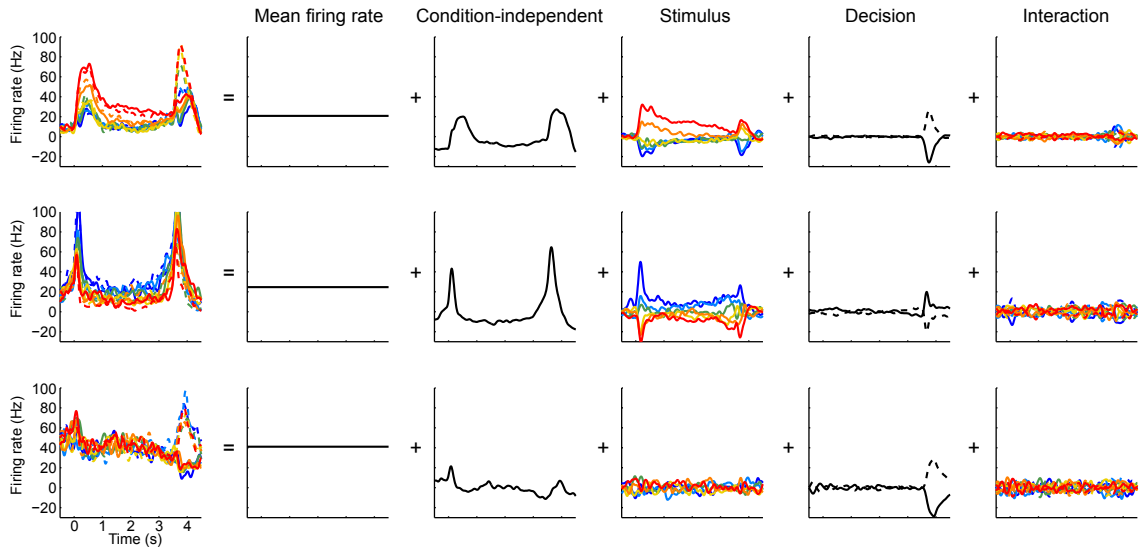

**Figure S11. Marginalization procedure.** PSTHs of three exemplary neurons from the somatosensory working memory task decomposed into marginalizations. See section S2.1 for details.

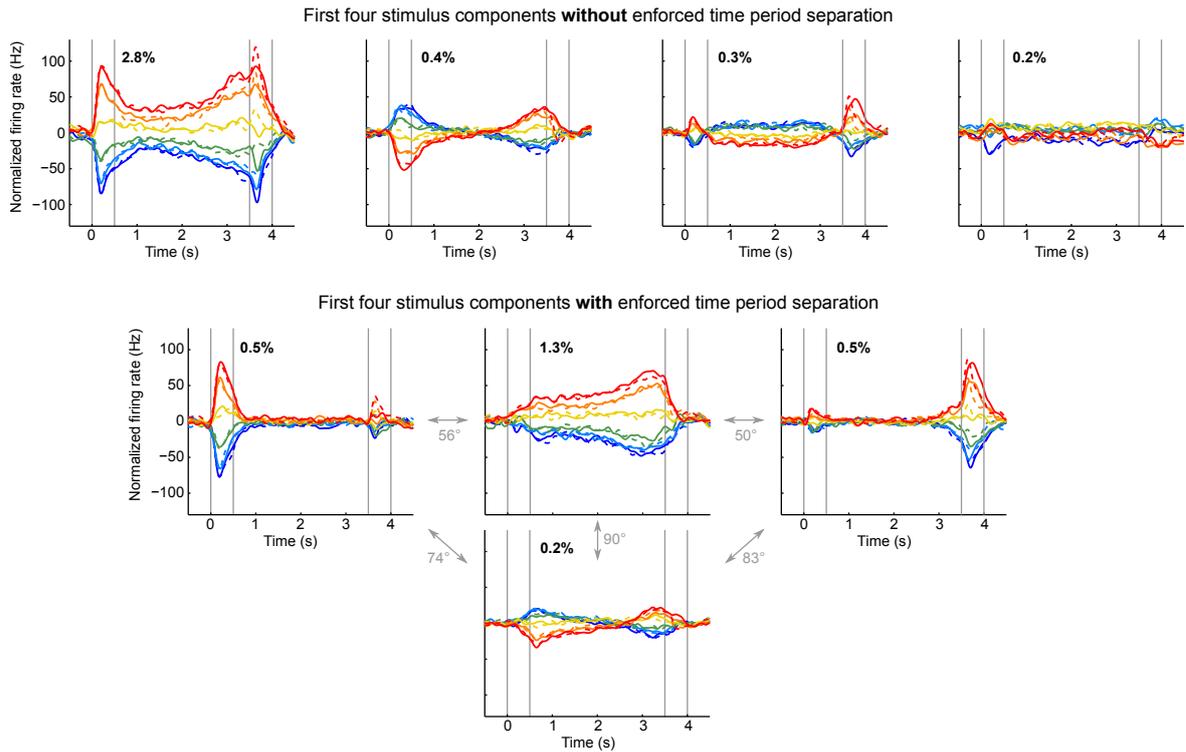

**Figure S12. Time period separation in dPCA.** On the top: first four stimulus components in the somatosensory working memory task, as obtained with dPCA without time period separation. On the bottom: first four stimulus components in the same dataset, with the additional time period separation constraint (see Methods for description). Components on the bottom are identical to the ones shown on Figure 2. The encoding axes for the components on the top are all mutually orthogonal, because they come from the same marginalization (see Methods), the stimulus one. However, the encoding axes for the components in the bottom are far from being orthogonal (angles are shown for various pairs of components). This is because now the algorithm effectively splits stimulus marginalization into three different ones (see Methods), corresponding to the three pre-defined time periods. Components shown here are arranged according to these time periods: first component has most of the variance in the S1 period, second and fourth — in the delay period, third – in the S2 period. Notice that the two components describing the delay period are exactly orthogonal, by construction. Proportion of variance of each projection is shown in each subplot in bold. It might seem that the components below account for substantially less variance than the components on the top, however the amount of *explained variance* (see Methods) by the first four components is very similar in both cases: 5.8% (top) and 5.4% (bottom).

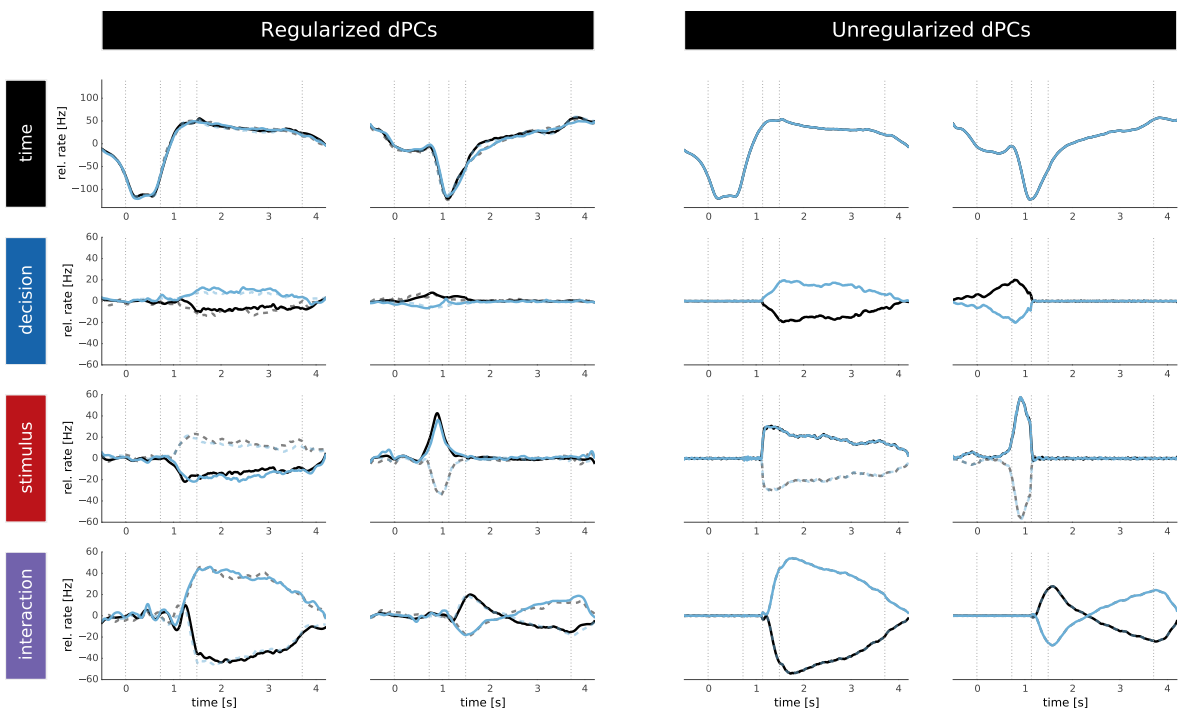

**Figure S13. Overfitting in dPCA.** Demixed principal components in the olfactory discrimination task in rats without re-stretching (two in each marginalization) in the standard regularized case (left) and without regularization, $\lambda = 0$ (right). The cross-validation error for un-regularized case is large, see Figure S14. See also section S2.3 for more details. The components on the left are the same ones presented in the main text in Figure 4.

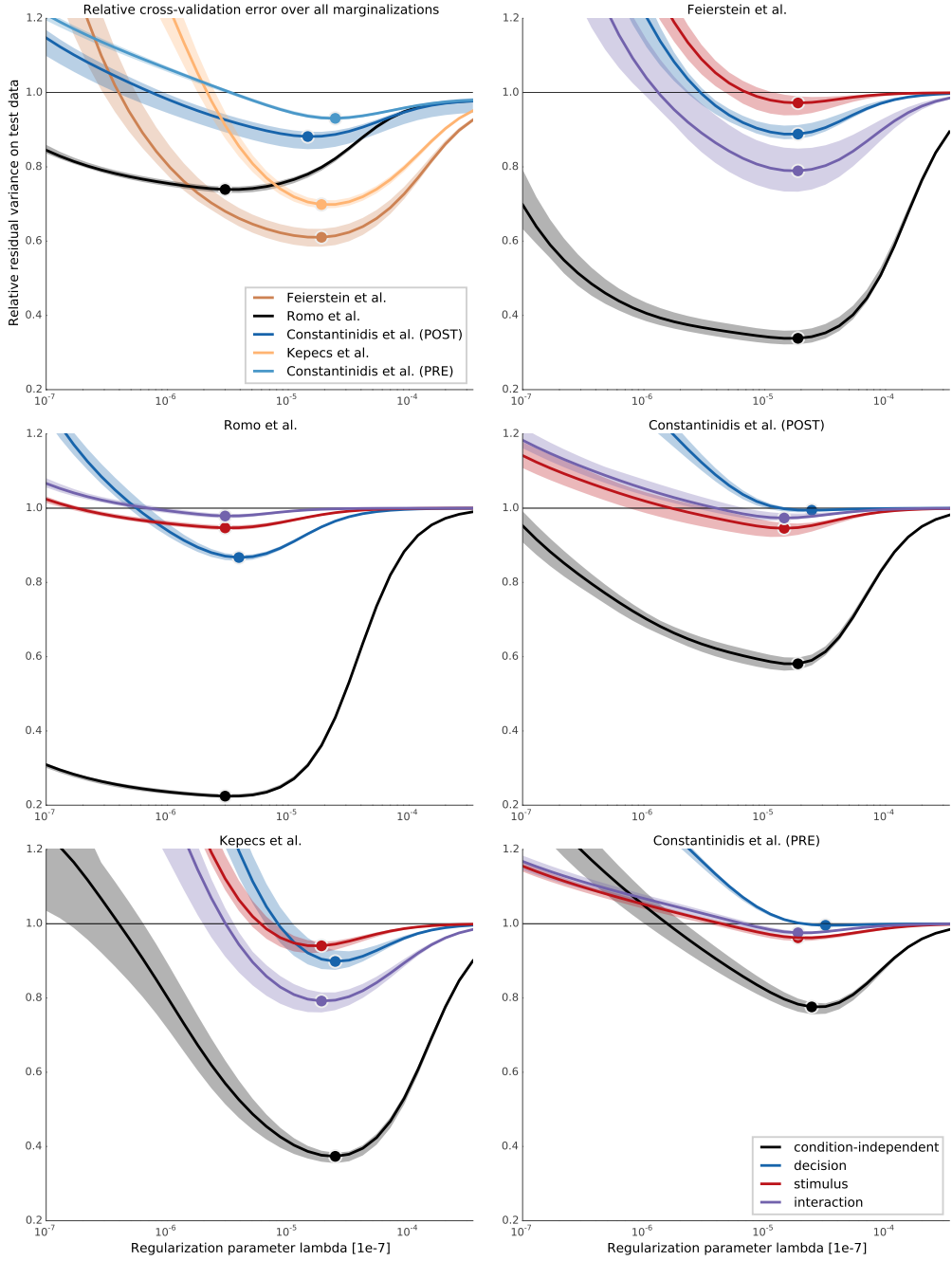

**Figure S14. Cross-validation errors depending on the regularization parameter $\lambda$.** Top left: mean (solid lines) and min/max (boundaries of shaded regions) of the relative cross-validation errors over all marginalizations for 10 repetitions. Different colors refer to different datasets, the minimum is marked by a dot. Other subplots: the same for each dataset, with different marginalizations showed separately. In each dataset the minimum cross-validation error is roughly the same across marginalizations, therefore we chose to use the same value of $\lambda$ in all marginalizations. See also section S2.3 for more details.

13

# S2 Supplementary notes

## S2.1 Marginalized averages

### S2.1.1 Intuition

In order to identify components that depend only on a small number of parameters, it is necessary to rigorously quantify the dependence of a component on the set of parameters or labels. For illustrative purposes, consider a function $x(t,s)$ that depends on time $t$ (ranging from 1 to T) and stimulus $s$ (ranging from 1 to S), Figure S15. We aim to segregate this function into contributions that are solely due to variations in either time or stimulus and contributions that are due to interactions of both parameters,

$$x(t,s) = \mu + z(t) + z(s) + z(t,s). \qquad (1)$$

There are many ways to perform this segregation. We pose, however, that the scalar $\mu$ should be the "best guess" about the value of $x(t,s)$ if neither time nor stimulus are determined. In other words, $\mu$ should simply be the *mean* of $x$,

$$\mu = \frac{1}{TS} \sum_{t,s} x(t,s) = \langle x(t,s) \rangle_{t,s} \qquad (2)$$

where $\langle . \rangle$ denotes the average over all subscripted parameters. The time-dependent term $z(t)$, on the other hand, should be the best estimate of the remainder $x(t,s) - \mu$ if the stimulus is undetermined. For a given time-point $t$ this is again just the average over all undetermined parameters,

$$z(t) = \langle x(t,s) - \mu \rangle_s . \qquad (3)$$

A similar reasoning applies to the stimulus component,

$$z(s) = \langle x(t,s) - \mu \rangle_t . \qquad (4)$$

Finally, the interaction component $z(t,s)$ is simply the remainder,

$$z(t,s) = x(t,s) - \mu - z(t) - z(s). \qquad (5)$$

We call each term of this specific segregation a *marginalized average* and note a range of useful properties: (1) Most trivially, the segregation is complete, i.e. we don't loose any information about the original function. (2) Averaging any marginalized average over one of its parameters gives zero, i.e.

$$\langle z(t) \rangle_t = 0, \qquad \langle z(s) \rangle_s = 0, \qquad (6)$$
$$\langle z(t,s) \rangle_t = 0, \qquad \langle z(t,s) \rangle_s = 0. \qquad (7)$$

This property is useful to see that, (3), the different terms are independent, i.e their covariance is zero, e.g.

$$\langle z(t) z(s) \rangle_{t,s} = 0 \qquad (8)$$

or

$$\langle z(t) z(t,s) \rangle_{t,s} = 0. \qquad (9)$$

In this case the variance of $x(t,s)$ can be written as the sum of the marginalized variance,

$$\mathrm{Var}(x(t,s)) = \mathrm{Var}(z(t)) + \mathrm{Var}(z(s)) + \mathrm{Var}(z(t,s)). \qquad (10)$$

This property finally allows a well-grounded statement on how the variance of $x(t,s)$ can be attributed to its parameters. We will make extensive use of this property when we define the objective for our demixing procedure. Before we generalize the notion of marginalized averages to vector-valued functions and more parameters, we clarify the relationship to a well-known statistical method.

### S2.1.2 Analysis of Variance

The marginalization procedure we introduced above turns out to be exactly equivalent to a linear model with categorical predictors used in ANOVA (analysis of variance) tests[1].

ANOVA (in this case, two-way ANOVA) is often applied as a statistical procedure to test whether a function like $x(t,s)$ varies with time, with stimulus, or with their interaction. In ANOVA terminology, time and stimulus are called *factors* and their individual values are called *levels*. If firing rate significantly varies with time, then we talk about significant *main effect* of time; the same goes for stimulus. If the effect of time is different for different stimulus values (or vice versa), then there is significant *interaction effect* between time and stimulus.

Usually ANOVA is applied when there are multiple measurements for each parameter combination. In our case there is exactly one value for each parameter combination (because we average over trials as a pre-processing step; this has to be done,

---

[1] See e.g. Christensen R, 2011, Plane answers to complex questions: the theory of linear models (Springer) or Rutherford A, 2001, Introducing ANOVA and ANCOVA: a GLM approach (Sage).

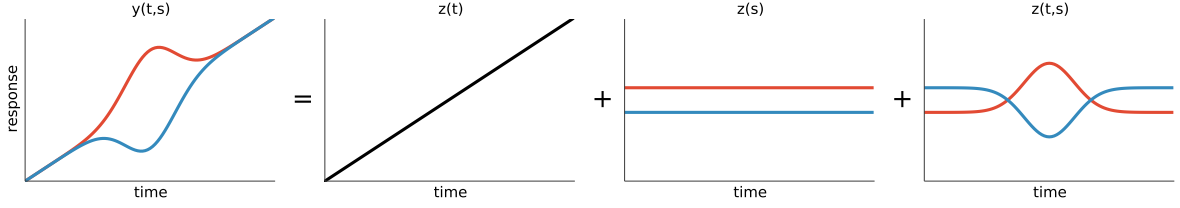

**Figure S15.** An illustrative segregation of a simple time- and stimulus-dependent function $x(t,s)$ into functions $z(t), z(s), z(t,s)$

because different neurons come from different sessions with different numbers of trials). In such a case ANOVA would have no or very little power to test for statistical significance of individual effects, but the underlying linear model still makes sense and is equivalent to our marginalization procedure, as we show below.

To formulate the linear model, we introduce categorical (dummy) variables $x_\tau, x_\sigma$ and $x_{\tau\sigma}$,

$$x_\tau = \begin{cases} 1, & \text{if } t = \tau \\ 0, & \text{otherwise.} \end{cases} \quad (11)$$

$$x_\sigma = \begin{cases} 1, & \text{if } s = \sigma \\ 0, & \text{otherwise.} \end{cases} \quad (12)$$

$$x_{\tau\sigma} = \begin{cases} 1, & \text{if } t = \tau \text{ and } s = \sigma \\ 0, & \text{otherwise.} \end{cases} \quad (13)$$

and consider the regression problem

$$x(t,s) = \mu + \sum_\tau z_\tau x_\tau + \sum_\sigma z_\sigma x_\sigma + \sum_{\tau,\sigma} z_{\tau\sigma} x_{\tau\sigma}. \quad (14)$$

This regression problem is under-determined since there are more unknown parameters $(1+T+S+TS)$ than data points $(TS)$. A standard set of constraints is therefore introduced in ANOVA as follows:

$$\sum_\tau z_\tau = 0, \quad (15)$$

$$\sum_\sigma z_\sigma = 0, \quad (16)$$

$$\forall \tau: \quad \sum_\sigma z_{\tau\sigma} = 0, \quad (17)$$

$$\forall \sigma: \quad \sum_\tau z_{\tau\sigma} = 0. \quad (18)$$

The solution to this regression problem is exactly equivalent to the marginalized averages $z_\tau \equiv \bar{x}(\tau), z_\sigma \equiv \bar{x}(\sigma)$ and $z_{\tau\sigma} \equiv \bar{x}(\tau,\sigma)$, see section S2.6, lemma 6.

### S2.1.3 Generalization

We seek to generalize the definition of marginalized averages along two lines: first, to more than one dependent variable and second, to more than two parameters. The first generalization is straightforward: We apply the marginalization procedure to every single dependent variable. This amounts to replacing all scalar variables in (1-5) with vector-valued quantities.

To generalize the procedure to more than two parameters we need to introduce a slightly more cumbersome notation. Instead of explicitly referencing every parameter (like time and stimulus), we only reference the set of parameters $S$. In the two-parameter case above, the set would be $S = \{t, s\}$. More generally, the number of parameters is $|S|$ and $f(S)$ denotes a function that depends on all parameters in $S$. Similarly, a function $f(\phi)$ depends only on a subset $\phi \subseteq S$ of parameters. This notation makes it possible to generalize the splitting (1) as

$$x(S) = \sum_{\phi \subseteq S} z(\phi) \quad (19)$$

where the sum goes over all possible subsets of parameters in $S$. In the two-parameter case the set of subsets is given by $\{\varnothing, \{t\}, \{s\}, \{t,s\}\}$ and we equated $z(\varnothing)$ with the mean $\mu$ of the data. The corresponding marginalized averages are,

$$\bar{x}(\phi) = \langle x(S) \rangle_{S \setminus \phi} - \sum_{\tau \subset \phi} \bar{x}(\tau), \quad (20)$$

where $\langle x(S) \rangle_{S \setminus \phi}$ denotes the averaging of $x(S)$ over all parameters $y \in S$ that are not also elements of $y \notin \phi$.

15

We already noted a range of useful properties of marginalized averages, all of which have a straight-forward generalization (see section S2.6 for proofs):

1. The average of $\bar{x}(\phi)$ over any parameter $y \in \phi$ is zero:
$$\langle \bar{x}(\phi) \rangle_y = 0 \quad \text{for} \quad y \in \phi. \quad (21)$$

   In other words, it is not possible to split the marginalized averages $z(\phi)$ into a sum of functions with reduced parameter dependence in any meaningful way.

2. They are independent, i.e. their covariance is zero:
$$\langle \bar{x}(\phi)\bar{x}(\phi') \rangle_S = 0. \quad (22)$$

   This, in turn, allows to segregate the data covariance matrix $\mathbf{C} = \langle \mathbf{x}(S)\mathbf{x}(S)^\top \rangle_S$ into a sum of marginalized covariances $\mathbf{C}_\phi = \langle \bar{\mathbf{x}}(\phi)\bar{\mathbf{x}}(\phi)^\top \rangle_S$,
$$\mathbf{C} = \sum_\phi \mathbf{C}_\phi. \quad (23)$$

3. There is an explicit formula for marginalized averages that does not rely on the hierarchical computation from small to larger numbers of parameters (i.e. is independent of the order of evaluation),
$$\bar{x}(\phi) = \sum_{\tau \subseteq \phi} (-1)^{|\tau|} \langle x \rangle_{(S/\phi) \cup \tau}. \quad (24)$$

### S2.2 dPCA optimization

Let $\mathbf{X}$ be a multidimensional data-matrix such that the first index refers to the observable (e.g. cell identity) whereas all subsequent indices refer to different parameters. As an example, $\mathbf{X}_{nts}$ refers to the mean response of neuron $n$ at time $t$ and $s$. For matrix products $\mathbf{Z} = \mathbf{U}\mathbf{X}$ we use the usual generalization,
$$Z_{kts} = \sum_n \mathrm{U}_{kn} X_{nts}. \quad (25)$$

Similarly for the $L_2$-norm,
$$\|\mathbf{X}\|_2^2 = \sum_{n,t,s} X_{nts}^2. \quad (26)$$

Note that for algorithmic implementations it is straight-forward to simply collapse all but the first coordinate into one axis, thereby embedding the multi-dimensional data-matrix into the standard two-dimensional matrix space.

The marginalized data is denoted by $\mathbf{X}_\phi$. Note that we assume that $\mathbf{X}_\phi$ has the same shape as the original data; parameters that have been averaged out during the marginalization are inserted back as replications (e.g. $\mathbf{X}_t$ varies only along the temporal axis but is constant in stimulus).

We can now define the objective for dPCA as a matrix equation,
$$L_{\text{dPCA}} = \sum_{\phi \subseteq S} \|\mathbf{X}_\phi - \mathbf{F}_\phi \mathbf{D}_\phi \mathbf{X}\|_2^2. \quad (27)$$

The objective is defined such that we project the full data $\mathbf{X}$ by means of $\mathbf{D}_\phi$ (the *decoder*) on a low-dimensional latent space (the dPCs) and then aim to reconstruct the marginalized data by means of a projection matrix $\mathbf{F}_\phi$ (the *encoder*). In this way we favour variance in a specified marginalization $\phi$ and punish variance from other marginalizations.

Note that we do not necessarily sum over all marginalizations but instead often pool marginalizations that cannot be independent. As an example, consider a behavioural parameter like decision that usually carries variance only at the end of a trial. Inevitably, there is no way that decision can be independent of time and so we add decision and decision-time together and consider them one "marginalization" ($\mathbf{X}_d + \mathbf{X}_{td} \to \mathbf{X}_{td}$).

One might be worried that the dPCA objective (27) is too biased towards demixed directions and might neglect directions with mixed variance even if these directions carry high amounts of variance. For this reason it is tempting to trade demixing against the total explained variance of the data,
$$\left\| \mathbf{X} - \sum_\phi \mathbf{F}_\phi \mathbf{D}_\phi \mathbf{X} \right\|_2^2. \quad (28)$$

It turns out, however, that in all neural datasets we analysed we never found that demixing alone leads to a strong decrease of the explained variance. Hence, all the data analysis in the subsequent sections will be based on dPCA objective (27) alone.

Minimizing objective (27) is straight-forward.

16

Replacing $\mathbf{M}_\phi = \mathbf{F}_\phi \mathbf{D}_\phi$ we can rewrite (27) as

$$\mathbf{M}_\phi = \arg\min_{\mathbf{M}} \|\mathbf{X}_\phi - \mathbf{MX}\|_2^2, \\ \text{s.t.} \quad r(\mathbf{M}) \leq s_\phi. \quad (29)$$

where $r(\mathbf{M})$ is the rank of $\mathbf{M}$ and $s_\phi$ is the number of components desired in marginalization $\phi$. Without a rank constraint the solution to this regression problem is simply given by $\mathbf{M}_\phi^* = \mathbf{X}_\phi \mathbf{X}^+$ where $\mathbf{X}^+$ denotes the pseudo-inverse. The rank-constrained problem ("reduced rank regression") is then solved by those $s_\phi$ projections that carry the most variance (see Reinsel et al. 1998 or Izenman 1975), i.e.

$$\mathbf{M}_\phi = \mathbf{U}_\phi \mathbf{U}_\phi^\top \mathbf{M}_\phi^* = \mathbf{U}_\phi \mathbf{U}_\phi^\top \mathbf{X}_\phi \mathbf{X}^+ \quad (30)$$

where $\mathbf{U}_\phi$ are the first $s_\phi$ left singular vectors of $\mathbf{M}_\phi^* \mathbf{X}$. Correspondingly, the encoder $\mathbf{F}_\phi = \mathbf{U}_\phi$ is given by the left singular vectors whereas the decoder is given by $\mathbf{D}_\phi = \mathbf{U}_\phi^\top \mathbf{X}_\phi \mathbf{X}^+$.

## S2.3 Regularization

Unfortunately, without any further constraints we run into serious over-fitting problems, see Figure S13. More precisely, if the number of neurons is on the same order as the number of independent data points (which is true for all datasets analysed here), then the distribution of eigenvalues of the data covariance matrix will peak at zero. In high-dimensional spaces, the space of eigenvectors with very small eigenvalues (the "kernel") will likely partially overlap with the leading PCs in individual marginalizations.

Without regularization, the decoder directions will end up lying within the kernel such that they overlap with the leading PC directions in one marginalization but are orthogonal to all other sources of variance. These directions will have very small variance, and the norms of decoder columns will be very large to compensate for that. With only slight instabilities of the kernel subspace, however, the same decoder will strongly amplify noise on a validation set, thereby leading to large cross-validation errors (see Figure S14).

A common approach to avoid such overfitting is to add a suitable regularization term. As we explained above, strong overfitting results in the large $L_2$-norm of the decoder. For that reason we add a regularization term to (29),

$$\mathbf{M}_\phi = \arg\min_{\mathbf{M}} \|\mathbf{X}_\phi - \mathbf{MX}\|_2^2 + \lambda \|\mathbf{M}\|_2^2, \quad (31)$$

subject to $r(\mathbf{M}) \leq s_\phi$, where $\lambda$ is a regularization parameter. Note that $\|\mathbf{M}\| = \|\mathbf{D}\|$, because

$$\|\mathbf{M}\|^2 = \|\mathbf{FD}\| = \text{Tr}(\mathbf{FDD}^\top \mathbf{F}^\top) \quad (32)$$
$$= \text{Tr}(\mathbf{F}^\top \mathbf{FDD}^\top) = \text{Tr}(\mathbf{DD}^\top) \quad (33)$$
$$= \|\mathbf{D}\|^2, \quad (34)$$

where we used orthogonality of $\mathbf{F}$.

We can rewrite this optimization problem in a more compact form that allows an analytic solution. Let $\mathbf{M} \in \mathbb{R}^{N \times N}$. We then stack $\mathbf{X}$ with the $N \times N$-dimensional identity matrix $\mathbf{I}_N$ and $\mathbf{X}_\phi$ with the $N \times N$-dimensional zero matrix $\mathbf{0}_N$ along the column index. As an example, if $\mathbf{X}_\phi \in \mathbb{R}^{N \times K}$, then the elements of the stacking are

$$[\mathbf{X}_\phi, \mathbf{0}_N]_{ij} = \begin{cases} [\mathbf{X}_\phi]_{ij} & \text{if } j \leq K, \\ [\mathbf{0}_N]_{i(j-K)} & \text{else.} \end{cases} \quad (35)$$

In this way we can rewrite the regularized objective (31) as

$$\mathbf{M}_\phi = \arg\min_{\mathbf{M}} \|\mathbf{X}_\phi - \mathbf{MX}\|_2^2 + \lambda \|\mathbf{M}\|_2^2, \quad (36)$$
$$= \arg\min_{\mathbf{M}} \|[\mathbf{X}_\phi, \mathbf{0}_N] - \mathbf{M}[\mathbf{X}, \lambda \mathbf{I}_N]\|_2^2. \quad (37)$$

subject to $r(\mathbf{M}) \leq s_\phi$. This formulation adheres to the same treatment as (29).

**Cross-validation to select $\lambda$**

We performed cross-validation in order to select the optimal value of the regularization parameter $\lambda$. To separate the data into training and testing sets, we held out one random trial for each neuron in each condition as a set of test "pseudo-trials" $\mathbf{X}_{\text{test}}$ (as the neurons were not recorded simultaneously, we do not have recordings of all $N$ neurons in any actual trial). Remaining trials were averaged to form a training set $\mathbf{X}_{\text{train}}$. Note that $\mathbf{X}_{\text{test}}$ and $\mathbf{X}_{\text{train}}$ have the same dimensions.

We then performed dPCA on $\mathbf{X}_{\text{train}}$ for various values of $\lambda$, selected 10 components in each marginalization (i.e. 40 components in total) to obtain $\mathbf{F}(\lambda)$ and $\mathbf{D}(\lambda)$, and computed reconstruction error $R(\lambda)$ for $\lambda$ between $10^{-7} \cdot \|\mathbf{X}\|^2$ and $10^{-3} \cdot \|\mathbf{X}\|^2$ (on a logarithmic grid). We repeated this procedure

10 times for different train-test splittings and averaged the resulting functions $R(\lambda)$. In all cases the average function $R(\lambda)$ had a clear minimum (Figure S13) that we selected as the optimal $\lambda$.

We used two approaches to compute $R(\lambda)$. The first approach uses $\mathbf{X}_\text{test}$ to predict $\mathbf{X}_\text{train}$:

$$R_1(\lambda) = \frac{\sum_\phi \|\mathbf{X}_{\text{train},\phi} - \mathbf{F}_\phi \mathbf{D}_\phi \mathbf{X}_\text{test}\|^2}{\|\mathbf{X}_\text{train}\|^2}, \quad (38)$$

where we wrote $\mathbf{F}(\lambda)$ and $\mathbf{D}(\lambda)$ as $\mathbf{F}$ and $\mathbf{D}$ for compactness. This is a fraction of training-set variance not explained by the test data. Note that it would not make sense to change $\mathbf{X}_\text{test}$ and $\mathbf{X}_\text{train}$ places here: the decoder and encoder are fitted to the training data, and should only be applied to the test data for the purposes of cross-validation.

The second approach uses $\mathbf{X}_\text{test}$ only. A naive implementation

$$\frac{\sum_\phi \|\mathbf{X}_{\text{test},\phi} - \mathbf{F}_\phi \mathbf{D}_\phi \mathbf{X}_\text{test}\|^2}{\|\mathbf{X}_\text{test}\|^2} \quad (39)$$

is incorrect: here we use $\mathbf{X}_\text{test}$ both for prediction and for evaluating this prediction, which is inacceptable (even though in practice often leads to very similar results). An alternative method predicts one dimension (i.e. one neuron) of $\mathbf{X}_\text{test}$ at a time, using all other dimensions:

$$\frac{\sum_\phi \sum_{n=1}^N \|[\mathbf{X}_{\text{test},\phi}]_n - [\mathbf{F}_\phi [\mathbf{D}_\phi]_{-n} [\mathbf{X}_\text{test}]_{-n}]_n\|^2}{\|\mathbf{X}_\text{test}\|^2},$$

where $[\cdot]_n$ stands for $n$-th coordinate and $[\cdot]_{-n}$ for all coordinates except of the $n$-th one. The inner sum in this expression can be computed analytically, resulting in

$$R_2(\lambda) =$$
$$\frac{\sum_\phi \|\mathbf{X}_{\text{test},\phi} - (\mathbf{F}_\phi \mathbf{D}_\phi - \text{diag}\{\mathbf{F}_\phi \mathbf{D}_\phi\}) \mathbf{X}_\text{test}\|^2}{\|\mathbf{X}_\text{test}\|^2}, \quad (40)$$

We verified that using $R_1(\lambda)$ and $R_2(\lambda)$ yielded the same optimal $\lambda$ in our data sets.

## S2.4 Relation to Linear Discriminant Analysis

A common dimensionality reduction technique related to dPCA is linear discriminant analysis (LDA). In this section we will argue that while LDA can be applied to neural data for the purposes of demixing, it is less suited than dPCA.

In the LDA ansatz, every data point $\mathbf{y}$ belongs to a class $k \in \{1, \ldots, K\}$. Data points within each class $k$ have a mean (centroid) $\boldsymbol{\mu}_k$ and covariance $\boldsymbol{\Sigma}_k$. LDA assumes that all classes have the same covariance, estimated[2] as $\boldsymbol{\Sigma}_W = \sum_k \boldsymbol{\Sigma}_k / K$ and called *within-class covariance*. On the other hand, the centroids of all classes form the *between-class covariance*

$$\boldsymbol{\Sigma}_B = \frac{1}{K} \sum_{k=1}^K (\boldsymbol{\mu}_k - \boldsymbol{\mu})(\boldsymbol{\mu}_k - \boldsymbol{\mu})^\top \quad (41)$$

where $\boldsymbol{\mu} = \sum_k \boldsymbol{\mu}_k / K$.

LDA aims at finding directions (discriminant axes) maximizing between-class and minimizing within-class variances. For the case of one axis $\boldsymbol{w}$, LDA maximizes the following function:

$$L_\text{LDA} = \frac{\boldsymbol{w}^\top \Sigma_B \boldsymbol{w}}{\boldsymbol{w}^\top \Sigma_W \boldsymbol{w}}. \quad (42)$$

The general LDA solution is to take eigenvectors of $\boldsymbol{\Sigma}_W^{-1} \boldsymbol{\Sigma}_B$ as discriminant axes (they will generally not be mutually orthogonal, but projections on them will have zero correlation).

Note that LDA is a *one-way* technique: there is only one parameter, or label, associated with each data point — namely, its class $k$. Also, LDA is traditionally applied in situations with only few classes and a large number of data points in each class. It is therefore neither straightforward, nor customary to apply LDA to a neural data set as the ones considered in this manuscript to find demixing projections. Nevertheless, it is possible to do, using the same marginalization approach developed here for dPCA.

Consider a data set with time $t$, stimulus $s$ and decision $d$ as labels of the data points (with total number of different labels being $T$, $S$, and $Q$ respectively). To find discriminant axes associated with e.g. stimulus, we can split the data into stimulus marginalization and everything else: $\mathbf{X} = \mathbf{X}_{st} + \mathbf{X}_{(\text{rest})}$. We can then treat all $SQT$ data points as belonging to one of the $ST$ classes (with

---

[2]Note that these formulas are given for the case when all classes have the same number of points; in case of non-balanced classes one has to use weighted averages.

centroids $\mathbf{X}_{st}$) and $Q$ points in each class. Consequently, $\mathbf{C}_{st}$ can be taken as between-class covariance matrix and $\mathbf{C} - \mathbf{C}_{st}$ as the within-class one.

For the case of e.g. the somatosensory working memory task ($T = 501$, $S = 6$, $Q = 2$) this would correspond to 3006 classes with 2 points per class. This is an unusual setting for LDA, but one can nevertheless proceed with applying LDA to find a set of discriminant axes. There is, however, by far not enough data points to reliably estimate within-class covariance matrix in the $N = 832$-dimensional space, and as LDA requires inverting this covariance matrix, it will result in severe over-fitting. A standard approach in such a case is to use regularized LDA (rLDA) by replacing $\mathbf{\Sigma}_W$ with a "shrinkage" estimator $(1 - \lambda)\mathbf{\Sigma}_W + \lambda \mathbf{I}$. We found that by using this approach and fine-tuning the regularization parameter $\lambda$, we can obtain components that are similar to the dPCA components (data not shown).

Indeed, the objective functions of dPCA and LDA (as introduced above) can be shown to be similar to each other. In the case of a single decoder $\mathbf{d}$ and encoder $\mathbf{p}$ the dPCA objective (27) becomes

$$\left\| \mathbf{X}_\phi - \mathbf{p}\mathbf{d}^\top \mathbf{X} \right\|_2^2 = \left\| \mathbf{X}_\phi - \mathbf{p}\mathbf{d}^\top \sum_\tau \mathbf{X}_\tau \right\|_2^2 \quad (43)$$

$$= \|\mathbf{X}_\phi\|_2^2 - 2\mathbf{p}^\top \mathbf{X}_\phi \mathbf{X}_\phi^\top \mathbf{d} + \sum_\tau \mathbf{d}^\top \mathbf{X}_\tau \mathbf{X}_\tau^\top \mathbf{d}, \quad (44)$$

$$= \|\mathbf{X}_\phi\|_2^2 + (\mathbf{d} - 2\mathbf{p})^\top \mathbf{\Sigma}_B \mathbf{d} + \mathbf{d}^\top \mathbf{\Sigma}_W \mathbf{d}, \quad (45)$$

where we used that the covariance between different marginalizations is zero (section S2.6, lemma 3) and we defined between- and within-class covariance matrices following the discussion above:

$$\mathbf{\Sigma}_B = \mathbf{X}_\phi \mathbf{X}_\phi^\top, \quad (46)$$

$$\mathbf{\Sigma}_W = \sum_{\tau \neq \phi} \mathbf{X}_\tau \mathbf{X}_\tau^\top. \quad (47)$$

To first order the encoder is equivalent to the first PC of $\mathbf{\Sigma}_B$ while the decoder $\mathbf{d} \approx \mathbf{p} + \mathbf{w}$ can be split such that $\mathbf{w}$ ensures the orthogonality to other sources of variance. Then $\mathbf{p}^\top \mathbf{\Sigma}_B \mathbf{d} \approx \mathbf{d}^\top \mathbf{\Sigma}_B \mathbf{d}$ and the dPCA objective function can be written as,

$$L_{\text{dPCA}} \approx \mathbf{d}^\top \mathbf{\Sigma}_B \mathbf{d} - \mathbf{d}^\top \mathbf{\Sigma}_W \mathbf{d}, \quad (48)$$

which corresponds to the LDA objective function given above (42) in that it aims at maximizing between-class and minimizing within-class variance.

Even though LDA approach as outlined above can result in components close to the ones found by dPCA, we consider dPCA a more principled and systematic approach to the demixing problem. In particular, LDA looks only for decoding axes (discriminant axes) and ignores the encoding axes. Therefore, unlike dPCA, it does not aim at representing the original data set, does not provide any useful measures of explained variance, etc.

There is a multi-way, or factorial, extension of LDA known as MANOVA (multivariate ANOVA). In MANOVA, each data point is labeled with several parameters, i.e. all points are split into classes in several different ways. Whereas LDA is usually understood as either a dimensionality reduction or classification technique, MANOVA is usually understood as a statistical test; however, the underlying mathematics is very similar. To use the same neural example, MANOVA asks if the neural population activity differs with time, stimulus, decision, or some interactions between these parameters (ANOVA asks exactly the same about a single neuron). To answer this question, MANOVA considers within-class covariance matrix $\mathbf{\Sigma}_W$ as a shared covariance matrix of all the smallest subclasses, e.g. groups of points with fixed time, stimulus, and decision values; it then proceeds to compute various between-class covariance matrices $\mathbf{\Sigma}_B$ for various parameters and to analyze $\mathbf{\Sigma}_W^{-1} \mathbf{\Sigma}_B$ in order to perform the statistical test (compare it with ANOVA that looks at the ratio of between-class and within-class variances, $B/W$, as the statisic of interest). Note, however, that in all cases considered in this manuscript the number of points with a given $(t, s, d)$ combination is exactly 1, so MANOVA cannot be applied. And even if it could be applied, it would not address the demixing: as we saw above, $\mathbf{\Sigma}_W$ should be defined differently for that purpose.

## S2.5 Fourier-like components from temporal variations

Consider the decision components in the somatosensory working memory task, Figure 2. Here the second and the third components are closely resembling the first and second temporal derivatives of the leading decision component. To illustrate why these components are likely to be artifacts of the underlying sampling process, consider a highly simpli-

fied example in which a population of $N$ neurons is encoding a one-dimensional bell-shaped signal $z(t)$ in the population vector $\mathbf{a}$, i.e the population response is given by $\mathbf{x}(t) = \mathbf{a}z(t)$. In this case the population response lies in the one-dimensional subspace spanned by $\mathbf{a}$ and covariance matrix has rank one:

$$C = \sum_t \mathbf{x}(t)\mathbf{x}^\top(t) = \mathbf{a}\mathbf{a}^\top \sum_t z(t)^2. \quad (49)$$

Now consider the case in which the neurons are not recorded simultaneously but are pooled from different sessions. In behavioural experiments it is unavoidable that the onset of neural responses will vary by tenths or hundreds of milliseconds. Hence, the individual response $y_i(t)$ of neuron $i$ will experience a small time-shift $\tau_i$:

$$x_i(t) = a_i z(t + \tau_i), \quad (50)$$

see Figure S3 for an illustrative example with Gaussian tuning curves.

If $\tau_i$ is small we can do a Taylor expansion around $t$:

$$y_i(t) = a_i z(t) + a_i \tau_i z'(t) + \mathcal{O}(\tau_i^2). \quad (51)$$

Note that for pedagogical purposes we neglect higher-order corrections, but the extension is straight-forward. Let $\boldsymbol{\tau}$ be the vector of time-shifts of the neurons and let $\mathbf{b} = \mathbf{a} \circ \boldsymbol{\tau}$ be the element-wise vector product of $\mathbf{a}$ and $\boldsymbol{\tau}$, i.e. $[\mathbf{a} \circ \boldsymbol{\tau}]_i = a_i \tau_i$. Then the population response can be written as

$$\mathbf{x}(t) \approx \mathbf{a}z(t) + \mathbf{b}z'(t). \quad (52)$$

Hence, the covariance matrix is approximately,

$$\mathbf{C} \approx \sum_t \mathbf{a}\mathbf{a}^\top z^2(t) + \mathbf{b}\mathbf{b}^\top z'^2(t) \quad (53)$$

where we assumed for simplicity that $\mathbf{a} \perp \mathbf{b}$. In other words, time-shifts between observations will result in additional PCA components that roughly resemble the temporal derivatives of the source component.

## S2.6 Mathematical proofs

**Lemma 1** (Completeness). *The segregation of $\boldsymbol{x}(S)$ into the marginals $\bar{\boldsymbol{x}}(\phi)$ is complete, i.e.*

$$\boldsymbol{x}(S) = \sum_{\phi \subseteq S} \bar{\boldsymbol{x}}(\phi). \quad (54)$$

The proof follows directly from the definition (20). In addition to being complete, the individual marginalized averages are not linearly dependent in any parameter,

**Lemma 2** (Linear parameter independence). *The average of $\bar{\boldsymbol{x}}(\phi)$ over any parameter $x \in \phi$ is zero,*

$$\langle \bar{\boldsymbol{x}}(\phi) \rangle_x = 0 \quad for \quad x \in \phi. \quad (55)$$

*Proof.* Proof by induction: For $y \in S$ the average over $y$ is zero,

$$\langle \bar{\mathbf{x}}(x) \rangle_y = \langle \mathbf{x}(S) \rangle_S - \boldsymbol{\mu} = 0. \quad (56)$$

Let (55) be shown for all $\phi \subset S$ with $\|\phi\| \leq N$. Then, for $\|\phi\| = N+1$ and $y \in \phi$,

$$\langle \bar{\mathbf{x}}(\phi) \rangle_y = \langle \mathbf{x}(S) \rangle_{S \setminus (\phi \setminus y)} - \sum_{\tau \subset \phi} \langle \bar{\mathbf{x}}(\tau) \rangle_y, \quad (57)$$

$$= \langle \mathbf{x}(S) \rangle_{S \setminus (\phi \setminus y)} - \sum_{\tau \subseteq (\phi \setminus y)} \bar{\mathbf{x}}(\tau) \quad (58)$$

$$= \langle \mathbf{x}(S) \rangle_{S \setminus (\phi \setminus y)} - \sum_{\tau \subset (\phi \setminus y)} \bar{\mathbf{x}}(\tau) - \bar{\mathbf{x}}(\phi \setminus y), \quad (59)$$

$$= 0. \quad (60)$$

The last equation follows from the definition of $\bar{\mathbf{x}}(\phi \setminus y)$. □

From lemma 1 and lemma 2 we can directly derive the independence of two marginalized averages.

**Lemma 3** (Independence). *Different marginalized averages are independent, i.e.*

$$\langle \bar{\boldsymbol{x}}(\phi)\bar{\boldsymbol{x}}(\phi') \rangle_S = 0. \quad (61)$$

*Proof.* W.l.o.g. let $\phi \geq \phi'$. For $\xi = \phi/\phi' = \varnothing \Rightarrow \phi = \phi'$ so there is nothing to show. For $\xi = \phi/\phi' \neq \varnothing$ we use the linearity of the average and property (55),

$$\langle \bar{\mathbf{x}}(\phi)\bar{\mathbf{x}}(\phi') \rangle_S = \left\langle \langle \bar{\mathbf{x}}(\phi)\bar{\mathbf{x}}(\phi') \rangle_\xi \right\rangle_{S/\xi}$$

$$= \left\langle \langle \bar{\mathbf{x}}(\phi) \rangle_\xi \bar{\mathbf{x}}(\phi') \right\rangle_{S/\xi}$$

$$= 0.$$

□

From the last lemma, together with (54), we can obtain an identity for the sum of covariance matrices.

**Lemma 4** (Covariance splitting). *The covariance matrices of the marginalized averages sum up to the full covariance matrix,*

$$C = \sum_{\phi \subseteq S} C_\phi \qquad (62)$$

*for $C_\phi = \langle \bar{\boldsymbol{x}}(\phi) \bar{\boldsymbol{x}}^\top(\phi) \rangle$.*

Finally, we introduce an equivalent definition of the marginalized averages that only relies on certain averages of **x** and is independent of the order of evaluation.

**Lemma 5** (Reformulation). *For a given function $\boldsymbol{x}(S)$ the set of marginalized averages*

$$\bar{\boldsymbol{x}}(\phi) = \sum_{\tau \subseteq \phi} (-1)^{|\tau|} \langle \boldsymbol{x} \rangle_{(S/\phi) \cup \tau}. \qquad (63)$$

*is equivalent to (20).*

The proof of this lemma is not straight-forward. We first prove that any splitting with property (54) and (55) is unique.

**Lemma 6.** *Any segregation $\{\bar{\boldsymbol{x}}(\phi) | \phi \subseteq S\}$ of $\boldsymbol{x}(S)$ that obeys the properties (54) and (55) is unique.*

*Proof.* Let there be two sets $\{\bar{\mathbf{x}}\} \neq \{\bar{\mathbf{x}}'\}$ that obey (54) and (55). Then,

$$\sum_{\phi \subseteq S} \bar{\mathbf{x}}(\phi) = \sum_{\phi \subseteq S} \bar{\mathbf{x}}'(\phi). \qquad (64)$$

Using (55) we can average over $S/y$ with $y \in S$ to find

$$\bar{\mathbf{x}}(y) = \bar{\mathbf{x}}'(y). \qquad (65)$$

Averaging over $S/\{y,z\}$ leads to

$$\bar{\mathbf{x}}(y) + \bar{\mathbf{x}}(z) + \bar{\mathbf{x}}(y,z) = \bar{\mathbf{x}}'(y) + \bar{\mathbf{x}}'(z) + \bar{\mathbf{x}}'(y,z), \qquad (66)$$

$$\Leftrightarrow \bar{\mathbf{x}}(y,z) = \bar{\mathbf{x}}'(y,z). \qquad (67)$$

More generally, averaging over $S \setminus \xi$ with $\xi \subseteq S$ and $y \in S \setminus \xi$ yields

$$\sum_{\phi \subseteq S \setminus \xi} \bar{\mathbf{x}}(\phi) + \bar{\mathbf{x}}(\phi \cup \{y\}) = \sum_{\phi \subseteq S \setminus \xi} \bar{\mathbf{x}}'(\phi) + \bar{\mathbf{x}}'(\phi \cup \{y\}), \qquad (68)$$

$$\Leftrightarrow \bar{\mathbf{x}}(\xi \cup \{y\}) = \bar{\mathbf{x}}'(\xi \cup \{y\}), \qquad (69)$$

by induction. Hence, $\{\bar{\mathbf{x}}\} = \{\bar{\mathbf{x}}'\}$. $\square$

This lemma considerably simplifies the proof of the closed-form definition of the marginalized averages, lemma 5.

*Proof.* According to lemma 6 we only need to show that the marginalized averages (63) have the property (54) and (55). Regarding completeness, we rewrite $\mathbf{x}(S)$,

$$\mathbf{x}(S) = \sum_{\phi \subseteq S} \bar{\mathbf{x}}(\phi) \qquad (70)$$

$$= \sum_{\phi \subseteq S} \sum_{\tau \subseteq \phi} (-1)^{|\tau|} \langle \mathbf{x} \rangle_{(S \setminus \phi) \cup \tau} \qquad (71)$$

$$= \sum_{\phi \subseteq S} \sum_{\tau \subseteq \phi} (-1)^{|\tau|} \langle \mathbf{x} \rangle_{(S \setminus (\phi/\tau))} \qquad (72)$$

$$= \sum_{\xi \subseteq S} \sum_{\tau \subseteq S \setminus \xi} (-1)^{|\tau|} \langle \mathbf{x} \rangle_{(S \setminus \xi)} \qquad (73)$$

$$= \sum_{\xi \subseteq S} \langle \mathbf{x} \rangle_{(S \setminus \xi)} \left[ \sum_{\tau \subseteq S \setminus \xi} (-1)^{|\tau|} \right] \qquad (74)$$

From the binomial theorem we can compute the weighted sum over subsets,

$$= \sum_{\xi \subseteq S} \langle \mathbf{x} \rangle_{(S \setminus \xi)} (1-1)^{|S \setminus \xi|} \qquad (75)$$

$$= \mathbf{x}(S). \qquad (76)$$

To proof the linear parameter independence (55), note that, for any $y \in \phi$, we can rewrite (63) as

$$\bar{\mathbf{x}}(\phi) = \sum_{\tau \subseteq \phi/y} (-1)^{|\tau|} \left( \langle \mathbf{x} \rangle_{(S \setminus (\phi \setminus \tau))} - \langle \mathbf{x} \rangle_{(S \setminus (\phi \setminus \tau)) \cup y} \right). \qquad (77)$$

Averaging over $y$ cancels all terms on the right and hence $\langle \bar{\mathbf{x}}(\phi) \rangle_y = 0$. $\square$



\end{document}